\documentclass[prd,onecolumn]{revtex4}
\pdfoutput=1
\usepackage{amssymb,latexsym}
\usepackage{amsmath,amsbsy,bbm}
\usepackage{ifpdf}
\usepackage{epsfig,bm}
\usepackage{graphicx,comment}
\usepackage{color}
\usepackage{soul}
\usepackage{mathtools}
\usepackage{comment}
\usepackage[normalem]{ulem}
\begin{document}

\title{A neural network study of the phase transitions of the two-dimensional antiferromagnetic $q$-state Potts models on the square lattice}
\author{Yuan-Heng Tseng}
\affiliation{Department of Physics, National Taiwan Normal University,
88, Sec.4, Ting-Chou Rd., Taipei 116, Taiwan}
\author{Fu-Jiun Jiang}
\email[]{fjjiang@ntnu.edu.tw}
\affiliation{Department of Physics, National Taiwan Normal University,
88, Sec.4, Ting-Chou Rd., Taipei 116, Taiwan}
%\vspace{-2cm}

\begin{abstract}
The critical phenomena of the two-dimensional antiferromagnetic $q$-state Potts model on the square lattice with 
$q=2,3,4$ are investigated using the techniques of neural networks (NN). In particular,
an unconventional supervised NN which is trained using no information about the physics of the considered systems is employed. In addition, conventional unsupervised autoencoders (AECs)
are used in our study as well. Remarkably, while the conventional AECs fail to uncover
the critical phenomena of the systems investigated here, our unconventional supervised NN
correctly identifies the critical behaviors of all three considered antiferromagnetic 
$q$-state models. The results obtained in this study suggest convincingly that the applicability of our unconventional supervised NN is broader than one anticipates. In particular, when a new system
is studied with our NN, it is likely that it is not necessary to conduct any training, and one only needs to examine whether an appropriate reduced representation of the original raw configurations
exists, so that the same already trained NN can be employed to explore the related phase transition efficiently.

\end{abstract}

\maketitle

\section{Introduction}%\vskip-0.3cm

Neural networks (NN), both supervised and unsupervised ones, are employed recently to detect
various phases of many models. The outcomes of these applications demonstrate the potential power of NN in studying critical phenomena. In particular, NNs can detect the critical points with high accuracy \cite{Oht16,Tan16,Car16,Nie16,Den17,Li18,Chn18,Don19,Zha19,Tan20.1,Ale20,Fuk21,Tol23}.
Introductions regarding these mentioned applications of NNs
can be found in Refs.~\cite{Meh19,Car19}

While the Hamiltonians of the two-dimensional (2D) $q$-state Potts models on the square lattice are simple,
they have very rich critical phenomena \cite{Wu82}. Hence the 2D $q$-state Potts models, both ferromagnetic and antiferromagnetic ones, 
have triggered many related theoretical and numerical investigations. 

The phase diagrams of the antiferromagnetic $q$-state Potts models on the square lattice, where
the nearest couplings $J > 0$, are challenge due to the fact that thier ground states are highly degenerated \cite{Wan89,Wan90,Fer99}. At the moment, the general consensus regarding the phase structure of the antiferromagnetic $q$-state Potts model on the square lattice is summarized as follows. For $q \ge 5$, the phase transition
from the antiferromagnetic phase to the disordered phase is always first order. For $q = 4$,
the model is disordered at any temperature including the zero temperature. Moreover,
the critical temperature $T_c$ for $q=3$ antiferromagnetic Potts model on the square lattice 
is $T_c = 0$. Finally, the critical theory of the 2D 2-state antiferromagnetic Potts model (on the square lattice) is identical to that of the 2D 2-state ferromagnetic Potts model which has critical temperature $T_c \sim 1.1346$, and is in the 2D Ising universality class \cite{Wu82}.

When NNs are used to study the critical phenomena, either supervised or unsupervised NNs can be considered. While the use of supervised NNs to study a phase transition requires prior knowledge of the targeted critical point \cite{Car16,Nie16}, this is not needed for the unsupervised NNs \cite{Chn18,Ale20}. As a result,
the unsupervised NNs are favored over the supervised NNs in practice when investigating an unknown critical phenomenon. It should be pointed out that for both the conventional supervised and the customary unsupervised NNs, real physical quantities such as the spin configurations are employed for the associated training. Finally, whenever a new system or a new box size is considered, one may need to train a new NN in order to study the critical behavior associated with that new system or that new box size.

Apart from the conventional NN approaches for exploring the critical phenomena, efficient
untypical NNs are considered for studying phase transitions as well. Specifically,
these unconvetional NNs are trained with two very simple artificially made configuraitons. With dedicated strategies of building the required configurations for the NN predictions, the unconventional
NNs constructed in \cite{Tse22,Pen22,Tse23,Tse231} successfully determine the critical points of the three-dimensional (3D) classical $O(3)$ model, the
2D generalized 3-state $XY$ model, the one-dimensional (1D) Bose-Hubbard model, the
2D $q$-state ferromagnetic Potts model on the square lattice, the 2D classical $XY$ model,
the 2D quantum $U(1)$ link model,
and the 2D classical 6- and 8-state clock models. In particular, evidence shows that the calculations of the critical points of the models described above can be achieved by just one (supervised) NN which is trained only once on 200 sites. Due to its simple architecture, several hundred to few thousand factor in calculation speed is gained when the unconventional (supervised) NN introduced above is used.

In this study, we extend the applicability of the specified unconventional supervised NN 
to the 2D antiferromagnetic $q$-state Potts model on the square lattice with $q=2,3,4$.
For comparison purpose, we also study the critical phenomena of these models using
conventional deep autoencoders including the one constructed in Ref.~\cite{Ale20}.

Remarkably, while the conventional deep autoencodes fail to conclude definitely the critical behaviors of the antiferromagnetic 3- and 4-state Potts model, our NN successfully leads to
correct determination of the critical phenomena of these two models. The results demonstrated
here suggest that the applicability of our unconventional NN for studying phase transitions
is quite broad, and is more efficient than other conventional NNs 
since the same already trained NN can be recycled to study the 
critical phenomena of many models.

The rest of the paper is organized as follows. After the introduction, in Sect.~II the considered models
and the related observables are described. Then in Sect.~III the numerical results including both that from the Monte Carlo and the NN are presented. In particular, we show clearly that the conventional deep autoencoders employed here fail to give conclusive answers to the critical behaviors of the 2D antiferromagnetic 3- and 4-state Potts models. Finally, we summarize our investigation in 
Sect.~4.

\section{The considered models and observables}

The Hamiltonian $H_{\text{Potts}}$ of the 2D antiferromagnetic $q$-state Potts model on the square lattice considered here has the following expression \cite{Wu82,Wan89,Wan90}
\begin{equation}
\beta H_{\text{Potts}} = \beta \sum_{\left< ij\right>} \delta_{\sigma_i,\sigma_j},
\label{eqn}
\end{equation}
where $\beta$ is the inverse temperature ($1/T$), $\delta$ refers to the Kronecker function, and the Potts variable
$\sigma_i$ at each site $i$ takes an integer value from $\{1,2,...,q-1,q\}$.

The staggered magnetization $m$ for the antiferromagnetic model introduced above is defined by
\begin{eqnarray}
m = \frac{1}{q}\sum_{i=1}^{q} |M_i|.
\end{eqnarray}
Here $M_i$ is given as
\begin{eqnarray}
M_i = \frac{2}{L^2}\sum_x (-1)^{x_1 +x_2}\delta_{\sigma_x,i}, 
\end{eqnarray}
where the summation is over all lattice sites $x$ and $L$ is the linear system size. 
In addition, the susceptibility $\chi$ expressed by
\begin{equation}
\chi = \frac{1}{L^2}\frac{1}{q}\sum_{n=1}^{q}M_i^2
\end{equation}
is calculated here.
Finally, the Binder cumulant $Q_2$, which is defined as
\begin{equation}
Q_2 = \frac{\langle m^2\rangle^2}{\langle m^4\rangle},
\end{equation} 
is investigated as well. 

In this study, we explore the phase transitions associated
with the models described above for $q=2,3,4$ using both 
the NN method and the Monte Carlo simulatons.

\section{The numerical results}

To carry out the studies described above, we performed a large-scale Monte Carlo calculation using the Swendsen-Wang-Kotecky algorithm \cite{Wan89,Wan90}. In particular, for every considered temperature $T$ and linear system size $L$, the associated
spin configurations are recorded. These configurations will be used in the NN investigation.

\subsection{The Monte Carlo outcomes}

\subsubsection{The Monte Carlo results for the 2D 2-state antiferromagnetic Potts model}

For a second order phase transition, the finite-size scaling ansatzes related to $\chi$ and $m$
are given by
\begin{eqnarray}
&& \chi L^{-\gamma/\nu} = f_1(tL^{1/\nu}),\\
&& mL^{\beta/\nu} = f_2(tL^{1/\nu}).
\end{eqnarray}
Here $t = (T-T_c)/T_c$, $\gamma$, $\beta$, and $\nu$ are the associated critical exponents.
In addition, $f_1$ and $f_2$ are smooth functions of $tL^{1/\nu}$.

Based on the finite-size scaling ansatzes introduced above, if data of 
$\chi L^{-\gamma/\nu}$ ($m L^{\beta/\nu}$) are considered as functions of $tL^{1/\nu}$ for various $L$, then a smooth data collapse curve should appear.

For the 2-state antiferromagnetic Potts model on the square lattice, it is established that
$T_c \sim 1.1346$, $\gamma/\nu = 7/4$ $\nu = 1$, and $\beta/\nu = 1/8$ \cite{Wu82}.

\begin{figure*}
	%\vskip-0.5cm
	\hbox{~~~~~~
		\includegraphics[width=0.45\textwidth]{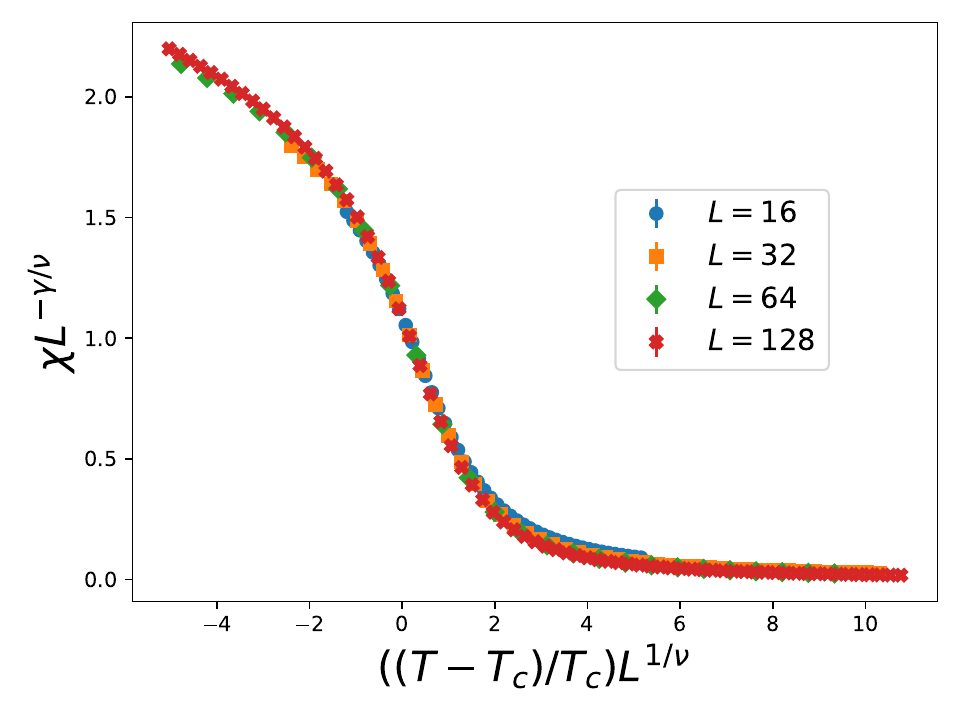}
		\includegraphics[width=0.45\textwidth]{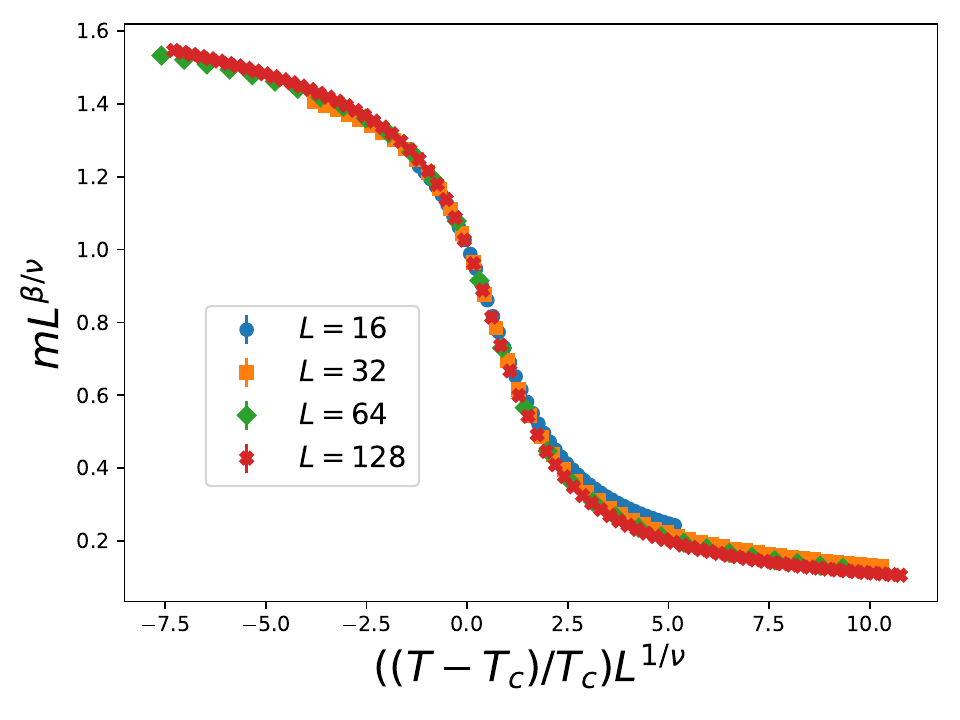}
	}
	%\vskip-0.2cm
	\caption{$\chi L^{-\gamma/\nu}$ v.s. $tL^{1/\nu}$ (left panel) and $mL^{\beta/\nu}$ v.s.
		$tL^{1/\nu}$ (right panel) for the 2-state antiferromagnetic Potts model on the square lattice. In obtaining the results, $\gamma/\nu = 7/4$, $\beta/\nu = 1/8$, $\nu=1$,
		and $T_c = 1.1346$ are used.		
	}
	\label{q2chimL}
\end{figure*}

With $T_c = 1.1346$, $\gamma/\nu = 7/4$, and $\beta/\nu = 1/8$, $\chi L^{-\gamma/\nu}$ v.s. $tL^{1/\nu}$ and  $mL^{\beta/\nu}$ v.s. $tL^{1/\nu}$ are shown as the left and the right panels
of fig.~\ref{q2chimL}. Smooth data collapse curves do appear in the figure. This suggests convincingly that our data are consistent with the 2D Ising universality class with the expected $T_c \sim 1.1346$, $\gamma/\nu = 7/4$, $\beta/\nu = 0.125$, and $\nu=1$.

\subsubsection{Monte Carlo results for the 2D 3-state antiferromagnetic Potts model}

\begin{figure*}
	%\vskip-0.5cm
	\hbox{~~~~~~
		\includegraphics[width=0.45\textwidth]{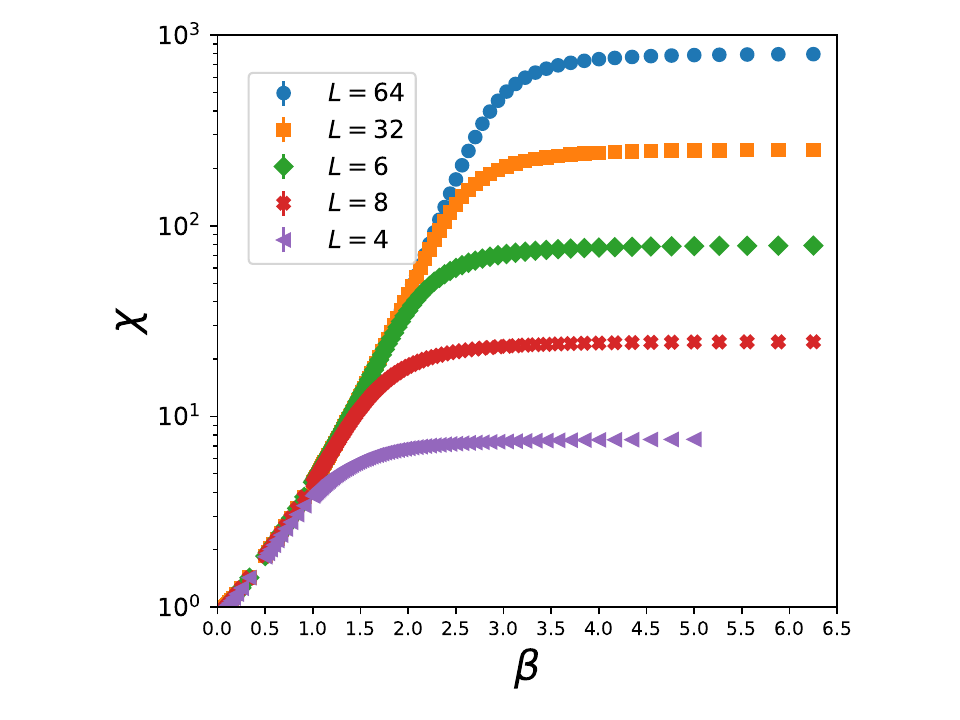}
		\includegraphics[width=0.45\textwidth]{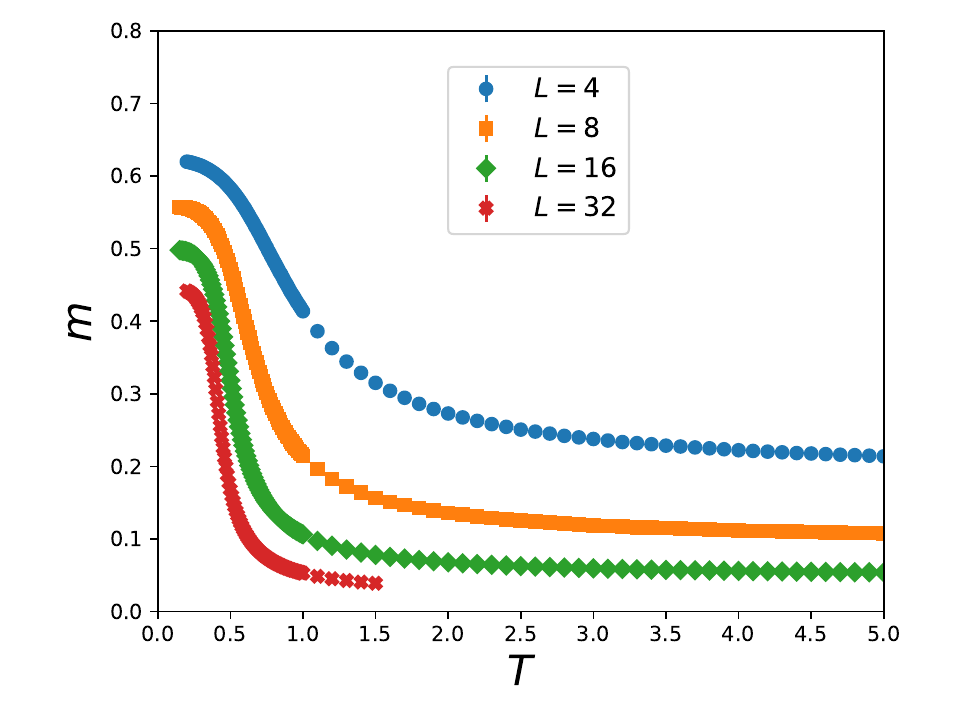}
	}
	%\vskip-0.2cm
	\caption{$\chi$ (left panel) and $m$ (right panel) as functions of $\beta$ and $T$ for several $L$ for the 3-state antiferromagnetic Potts model on the square lattice.		
	}
	\label{q3chimag}
\end{figure*}

Fig.~\ref{q3chimag} shows the $\chi$ (left panel) and $m$ (right panel) as functions of $\beta$ (or $T$) for several $L$.
The outcomes of the figure match quantitatively with that available in Ref.~\cite{Wan90}. This confirms
the correctness of our code(s) used for the Monte Carlo simulations.

It is well-known that the phase transition for 2D 3-state antiferromagnetic Potts model on the square lattice takes place at zero temperature \cite{Wan89,Fer99}. 

\begin{figure*}
	%\vskip-0.5cm
	\hbox{~~~~~~
		\includegraphics[width=0.475\textwidth]{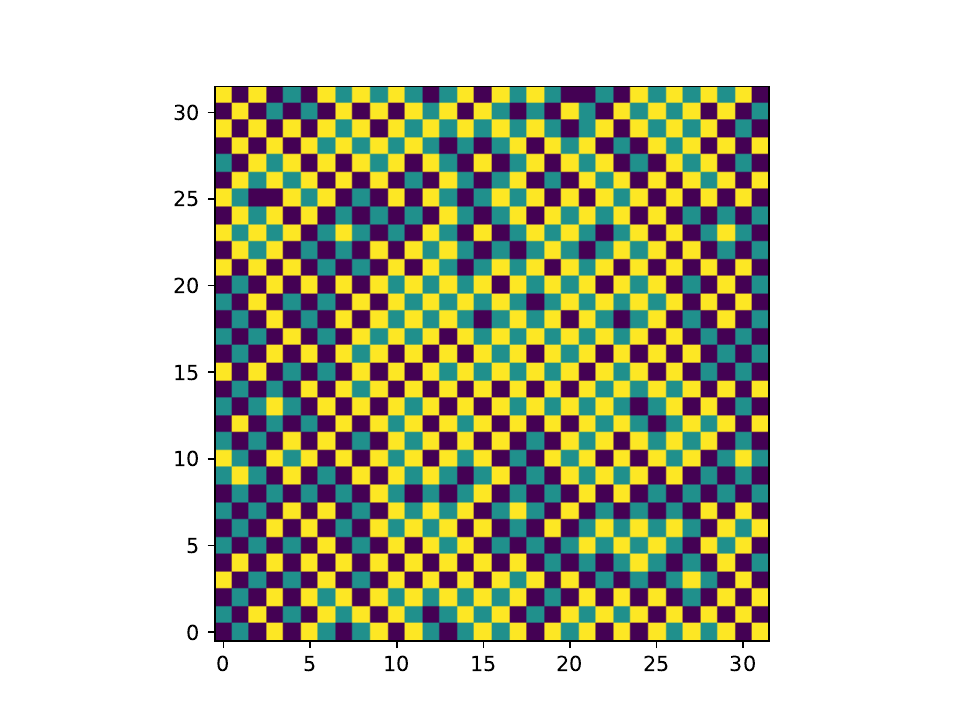}
		\includegraphics[width=0.475\textwidth]{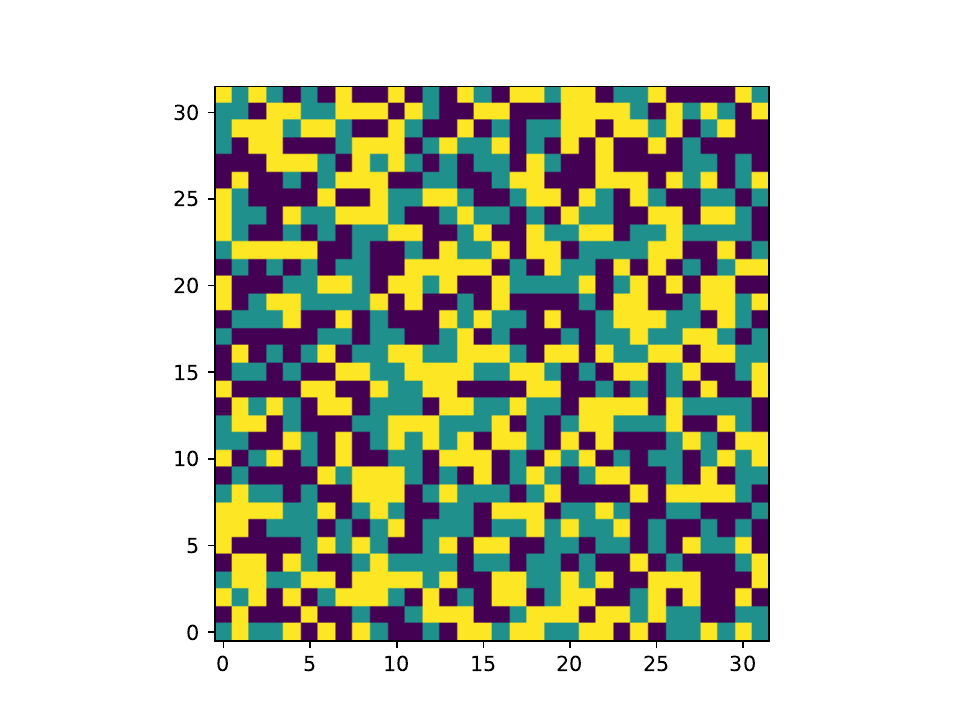}
	}
	%\vskip-0.2cm
	\caption{Snapshots of the spin configurations for the 3-state antiferromagnetic Potts model at $T=0.2$ (left panel) and at $T=10.0$ (right panel). The linear system sizes $L$ for both panels are $L=32$.		
	}
	\label{q3MCsnap32}
\end{figure*}

\begin{figure*}
	%\vskip-0.5cm
	\hbox{~~~~~~
		\includegraphics[width=0.475\textwidth]{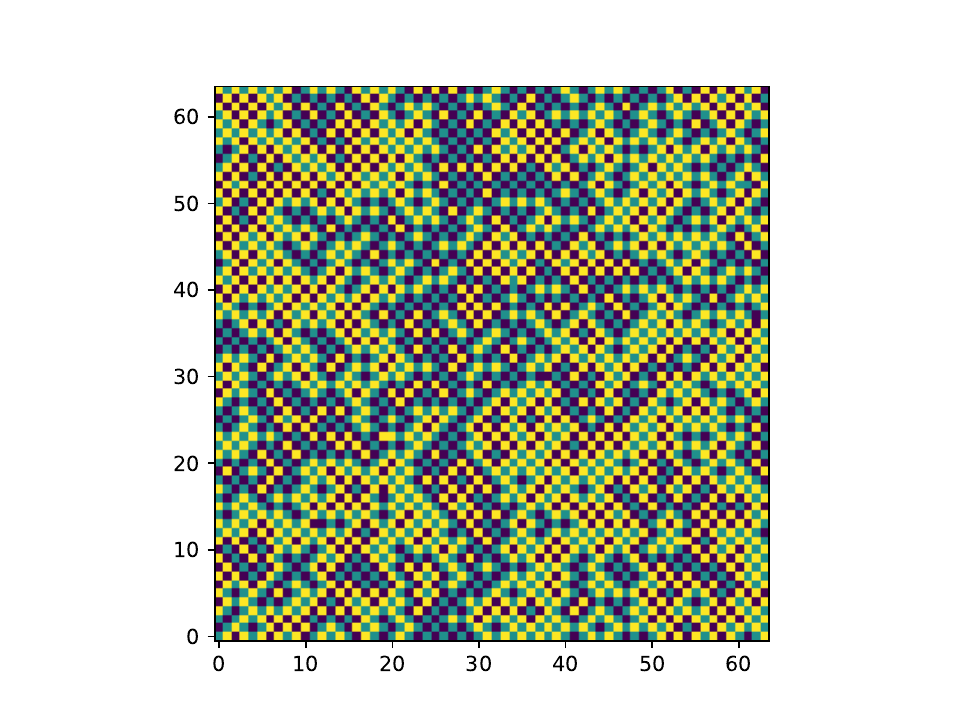}
		\includegraphics[width=0.475\textwidth]{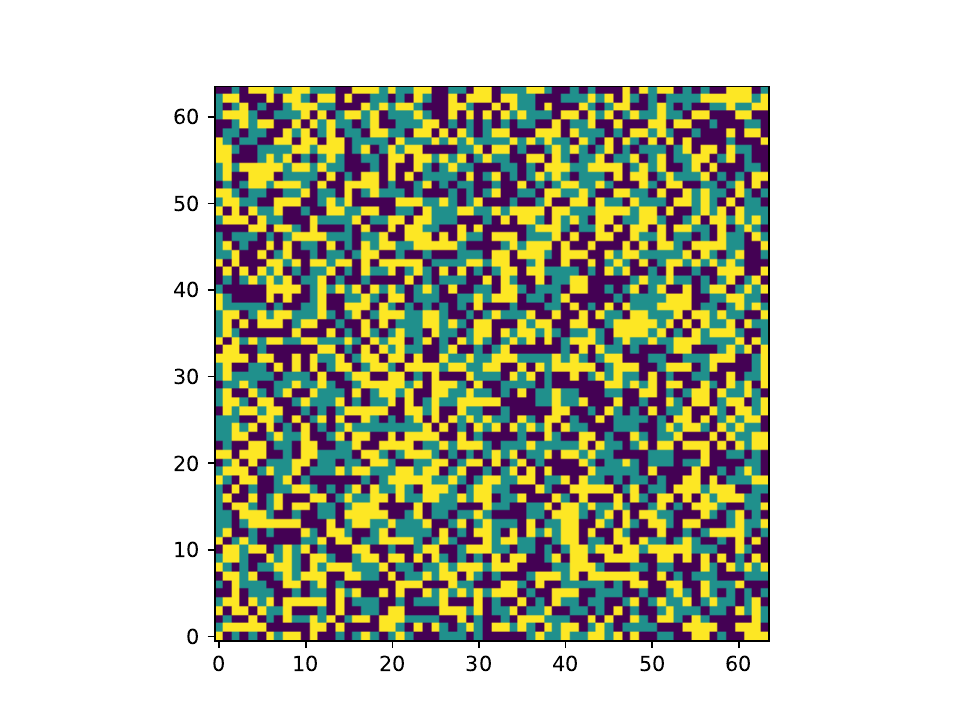}
	}
	%\vskip-0.2cm
	\caption{Snapshots of the spin configurations for the 3-state antiferromagnetic Potts model for $T=0.2$ (left panel) and $T=10.0$ (right panel). The linear system sizes $L$ for both panels are $L=64$.		
	}
	\label{q3MCsnap64}
\end{figure*}

Fig.~\ref{q3MCsnap32} shows the snapshots of the spin configurations for $T = 0.2$ (left panel) and $T=10.0$ (right panel). The linear system sizes for both panels of the figure are $L=32$.
It is interesting to notice that although $T_c = 0$ for this model, for finite $L$,
the feature of antiferromagnetic order in the spin configurations already appears at some low temperatures, such as $T=0.2$. 
This result also occurs for $L=64$ at $T=0.2$, sees fig.~\ref{q3MCsnap64}. 

One may argue that the appearance of antiferromagnetic order for $T \le 0.2$ is a finite-size
effect. In the following we will provide evidence to support the fact that such a behavior is unlikely a finite-size effect.

\begin{figure*}
	%\vskip-0.5cm
	\hbox{~~~~~~
		\includegraphics[width=0.45\textwidth]{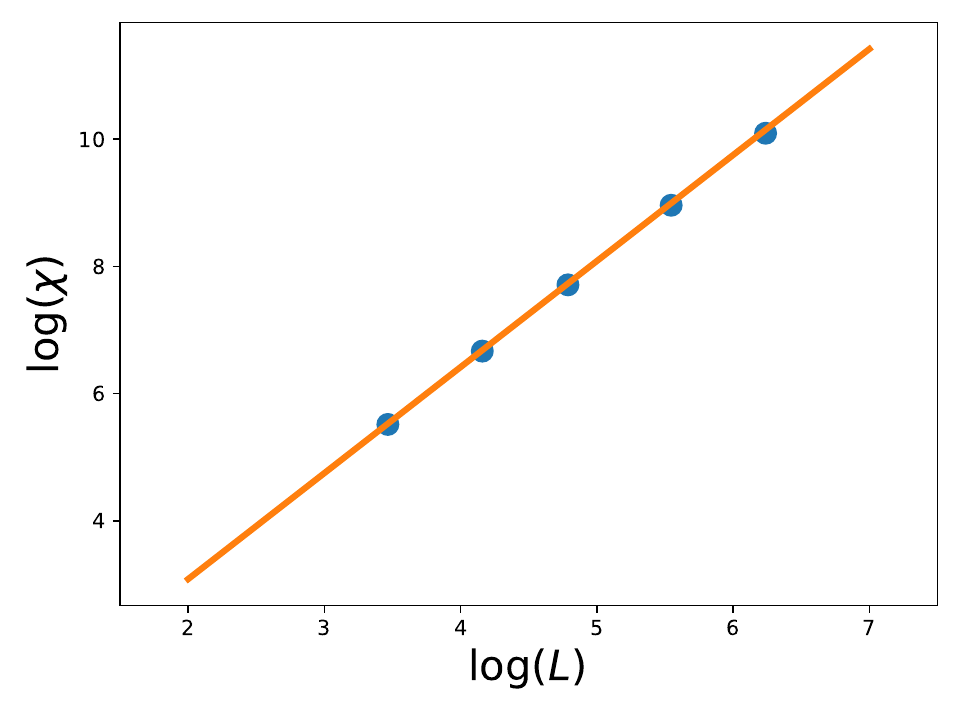}
		\includegraphics[width=0.45\textwidth]{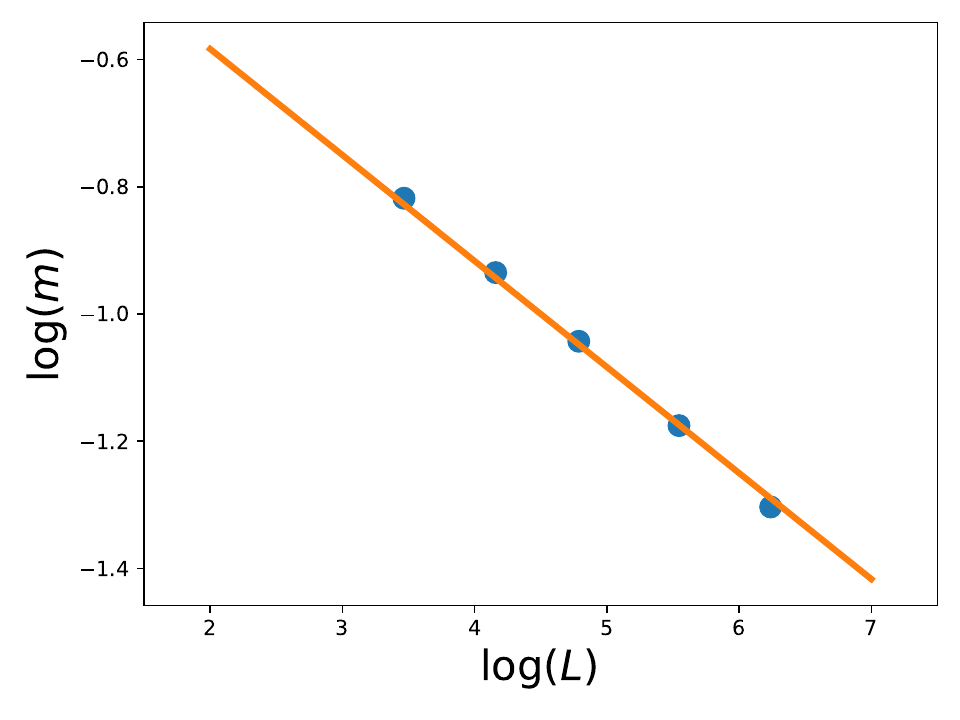}
	}
	%\vskip-0.2cm
	\caption{$\log(\chi)$ v.s $\log{L}$ (left panel) and $\log(m)$ v.s. $\log(L)$ (right panel ) for the 3-state antiferromagnetic Potts model. The data are obtained at  $T=0.2$.		
	}
	\label{q3logchim}
\end{figure*}

In the critical region, one expects $\chi \sim L^{\gamma/\nu}$ and $m \sim L^{-\beta/\nu}$.
For 3-state antiferromagnetic Potts model, $\gamma/\nu = 5/3$ and $\beta/\nu = 1/6$.
The left and the right panels of fig.~\ref{q3logchim} are $\log{\chi}$ v.s. $\log{L}$ and
$\log{m}$ v.s. $\log{L}$. The data for the figure are obtained at $T=0.2$. The solid lines
in the left and the right panels of fig.~\ref{q3logchim} have slopes of 5/3 and -1/6, respectively. With this one arrives at $\gamma/\nu = 5/3$ and $\beta/\nu = -1/6$. 
In other words, the data determined at $T=0.2$ are already in the critical region. 
Hence the antiferromagnetic ordering shows up in the left panel of fig.~\ref{q3MCsnap32}
is indeed a signal which supports the fact that the system is in the critical region.

\subsubsection{Monte Carlo results for the 2D 4-state antiferromagnetic Potts model}

As we have demonstrated, for the 3-state antiferromagnetic Potts model, one has $\chi \sim L^{\gamma/\nu}$ at $T_c$ (which is 0 theoretically for this model). In other words, $\chi$ scales with some power in $L$ near the critical region.

\begin{figure*}
	%\vskip-0.5cm
	\includegraphics[width=0.65\textwidth]{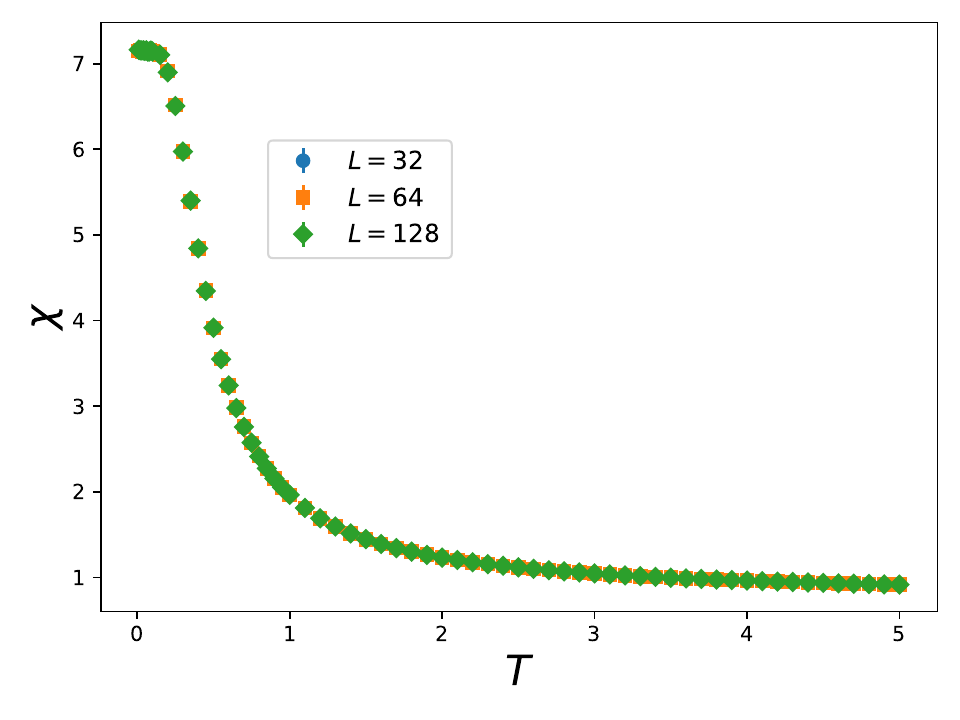}
	\vskip-0.2cm
	\caption{$\chi$ as functions of $T$ for $L=32,64,128$ for the 4-state antiferromagnetic Potts model on the square lattice.}
	\label{q4chi}
\end{figure*}

$\chi$ as functions of $T$ for $L=32,64,128$ for the 4-state antiferromagnetic Potts model
on the square lattice are shown in fig.~\ref{q4chi}. The saturation of $\chi$ to a constant (with a not large magnitude)
for all $L$ and all the considered values of $T$ indicates that the 4-state antiferromangetic
Potts model is disordered at all temperatures including the zero temperature. This observation
is consistent with the theoretical expectation \cite{Fer99}.

\subsection{The NN results}

\subsubsection{The employed MLP}

\begin{figure*}
  %\vskip-0.5cm
       \includegraphics[width=0.8\textwidth]{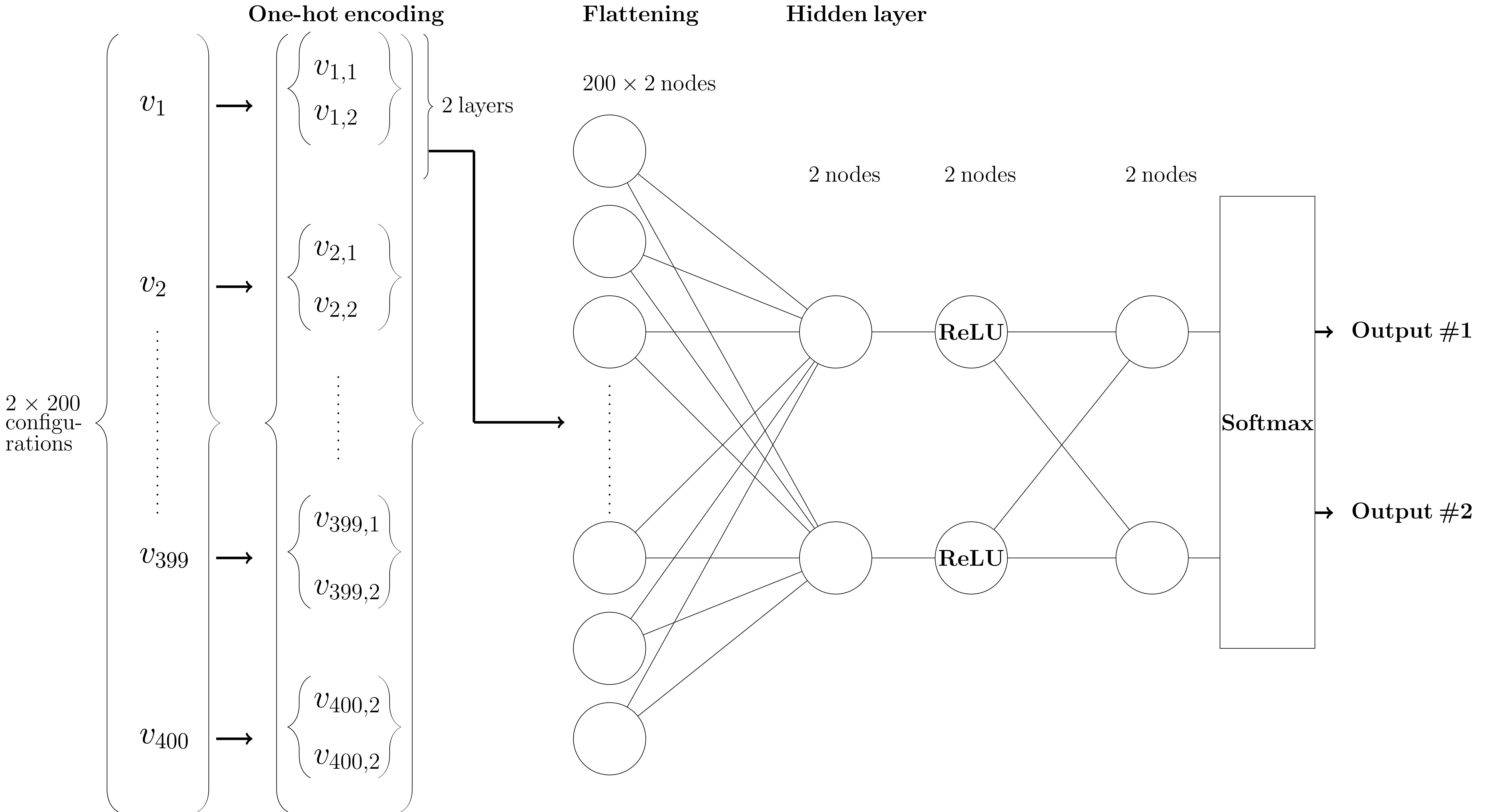}
        \vskip-0.2cm
        \caption{The MLP used in this study. The figure is adopted from Refs.~\cite{Pen22,Tse23,Tse231} since the same NN is used here.}
        \label{figMLP}
\end{figure*}

For this paper to be self-contained, in this subsubsection we described the considered NN briefly. Similar paragraphs have been appeared in some previous publications (including but not limited to Refs.~\cite{Pen22,Tse23,Tse231}). The NN used here is in principle the same as that employed in Refs.~\cite{Pen22,Tse23,Tse231}
which is built using keros and tensorflow \cite{keras,tens}. 
In summary, the employed NN is a multilayer perceptron (MLP) which consists of one input layer, one hidden layer of two neurons,
and one output layer. 
minibatch and adam are used as the main algorithm and optimizer of our NN. One-hot encoding, flattening, and $L_2$ regularization (which is used to avoid overfitting) are considered in the
NN architecture as well. The activations functions in the NN infrastructure are ReLU (softmax) in the hidden (output) layer. We have carried out 800 epochs
with a batch size of 40. The loss function used in the NN calculations is the categorical cross-entropy. Fig.~\ref{figMLP} is the typical graphical representation of our NN.

\begin{figure*}
	%\vskip-0.5cm
	\includegraphics[width=0.35\textwidth]{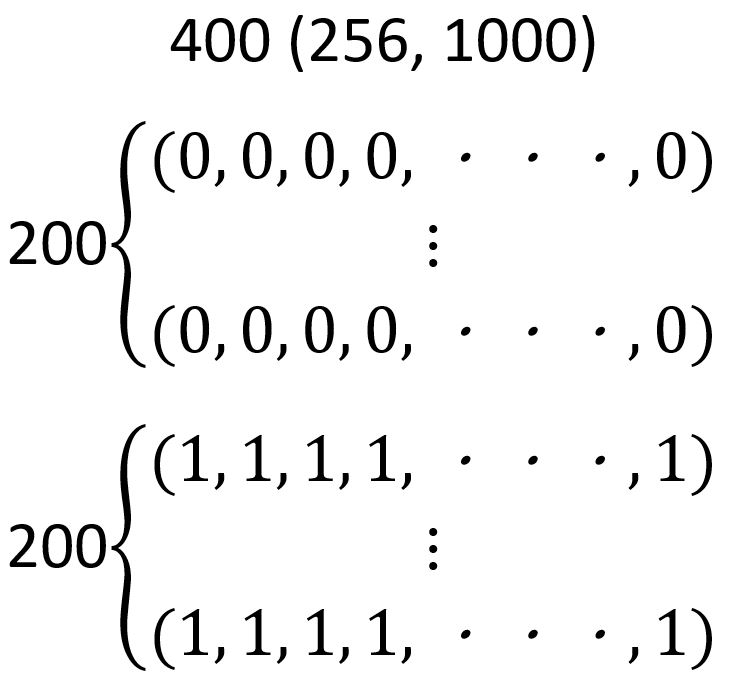}
	\vskip-0.2cm
	\caption{The trainig set used in this study. $256$ is employed for the calculations
	associated with $L=16$. A similar graph has shown up in Ref.~\cite{Jia24}.}
	\label{training}
\end{figure*}

The training set for the NN consists of 200 copies of two artificially made
configurations. Every configuration in the training set has 400 (1000) elements for the $q=2$ ($q=3,4$) model(s). In particular,
all the elements of the first and the second kinds of configurations have the values of 1 and 0, respectively. The labels used are automatically (1,0) and (0,1) since the training set has
two various kinds of configurations. The graphical representation of the training set is shown in fig.~\ref{training}.

Regaring the configurations used for the prediction, for $q=2,3,4,$ and for each of the generated configurations from the MC simulations, the first four hundred spins are employed
to construct the required configurations to feed the NN. For few cases, 1000 spins are used.

For $q=2,4$ models, spin values mod $2$ are assigned to the associated spots of the built configuration for the NN prediction. 

For $q=3$ model, if the spin value is 1 (2), then
1 (0) is assigned. In addition, if the spin value is 3, then 1 and 0 are assigned to the
related position of the configuration with equal probabilities.

\subsection{The NN results obtained from the MLP}

The standard deviations STD of the magnitude $R$ of the NN output vectors are used as 
the quantity to detect the critical point.

\subsubsection{The NN results of the 2D 2-state antiferromagnetic Potts model}

\begin{figure*}
%\vskip-0.5cm
\hbox{~~~~~~~
\includegraphics[width=0.425\textwidth]{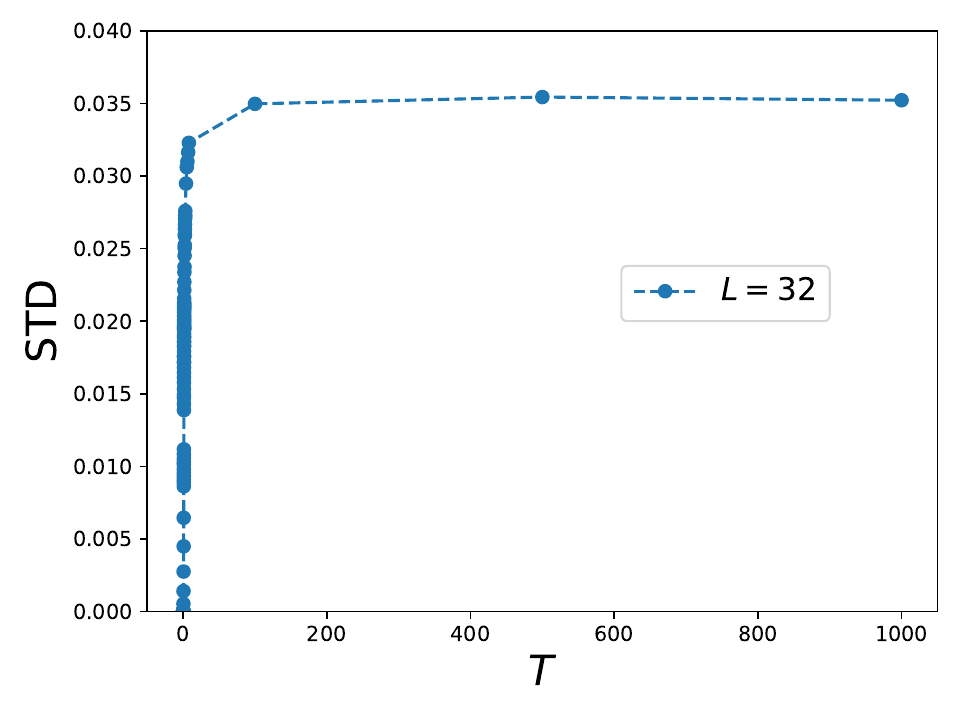}
\includegraphics[width=0.425\textwidth]{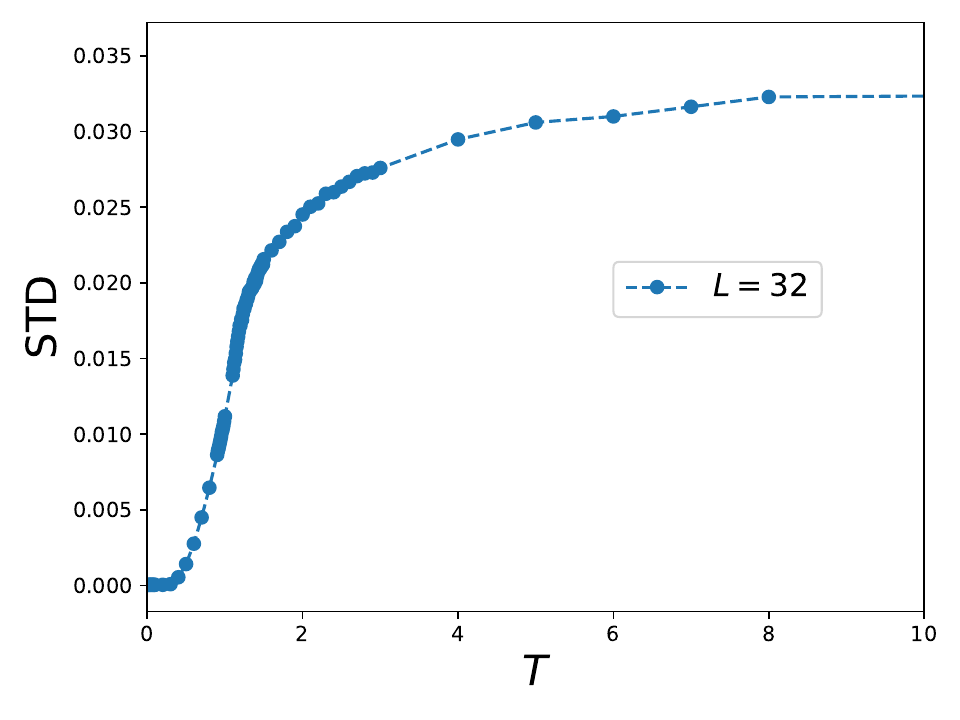}
}
%\vskip-0.2cm
\caption{(Left) STD as a function of $T$ for $L=32$ for the 2-state antiferromagnetic Potts model. (Right) The zoom in of the left panel in the temperature regin (0,10).}
\label{q2STDL32}
\end{figure*}

The STD of $R$ ($L=32$) as a function of $T$ for the 2-state antiferromagnetic Potts model
is shown in fig.~\ref{q2STDL32}. The figure implies that as one moves from the extremely low-temperature region to the high-temperature region, the values of STD increase monotonically from 0 to a constant $R_{\text{H}}$ (The value of this constant $R_{\text{H}}$ depends on the number of sites used for the training). Such an outcome indicates that at very low temperatures,
The difference between the values of $R$ of any two configurations is very tiny. In addition,
the mentioned difference amplifies with increasing $T$ untill it reaches a maximum degree. In
other words, as one goes from the low-temperature region to the high temperature region,
the configurations change from (extremely) high " similarity " to a little bit low " similarity ". 
This strongly suggests that there is a phase transition between the low-temperature region
and the high-temperature region.

For a given $L$, we define the corresponding pseudo-critical point $T_c(L)$ as the temperature where  
the $R$ curve (when they are considered as a function of $T$) and $R_{\text{H}}/2$ intersect. 

\begin{figure*}
	%\vskip-0.5cm
	\hbox{~~~~~~
		\includegraphics[width=0.31\textwidth]{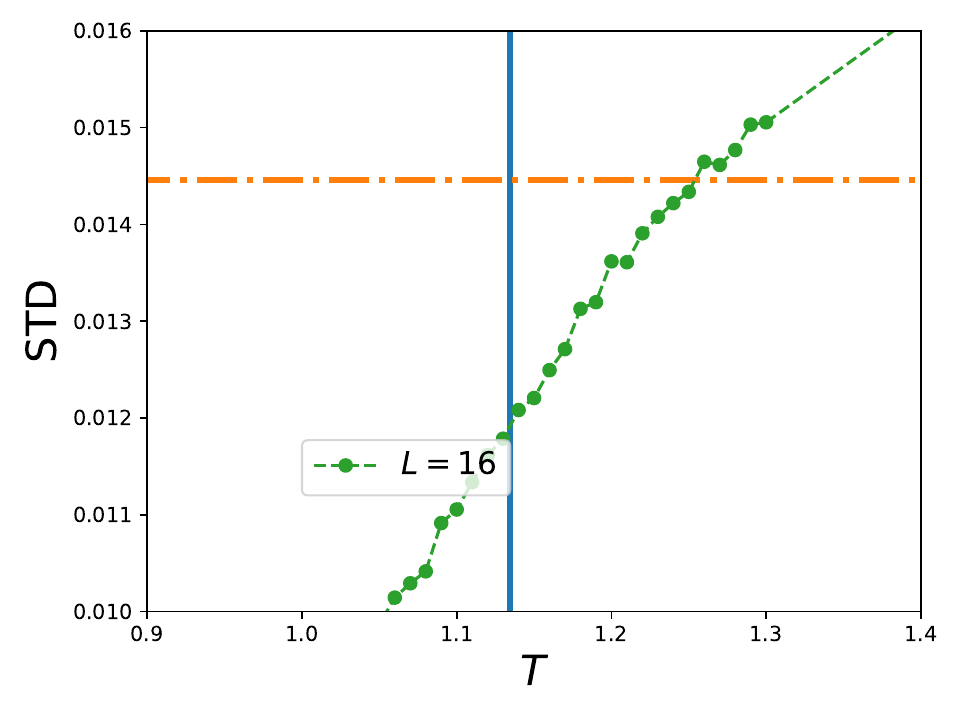}
		\includegraphics[width=0.31\textwidth]{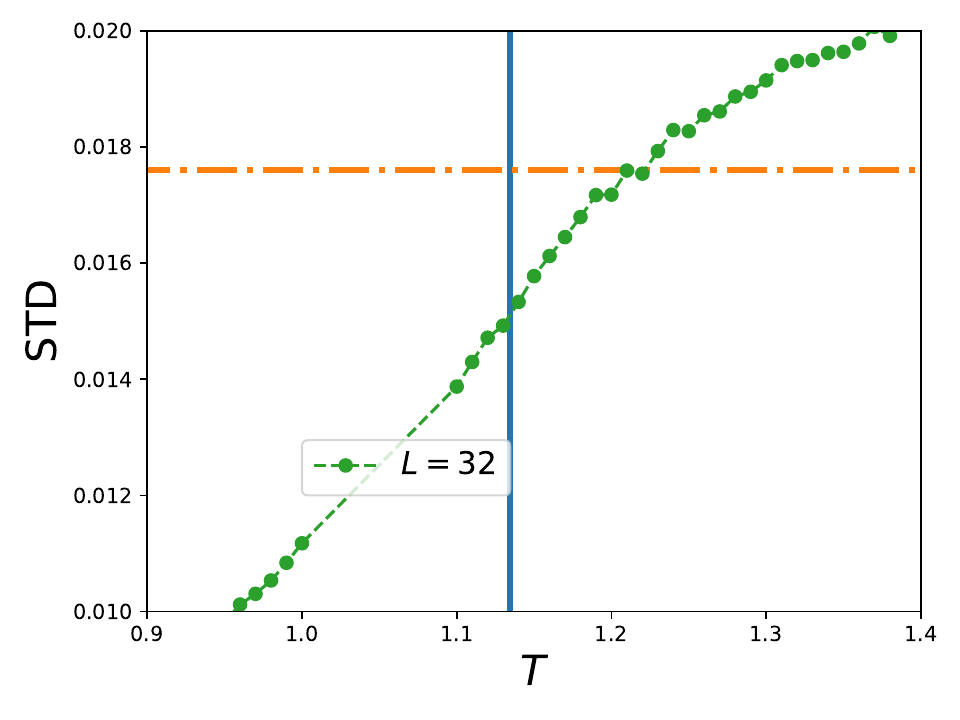}
		\includegraphics[width=0.31\textwidth]{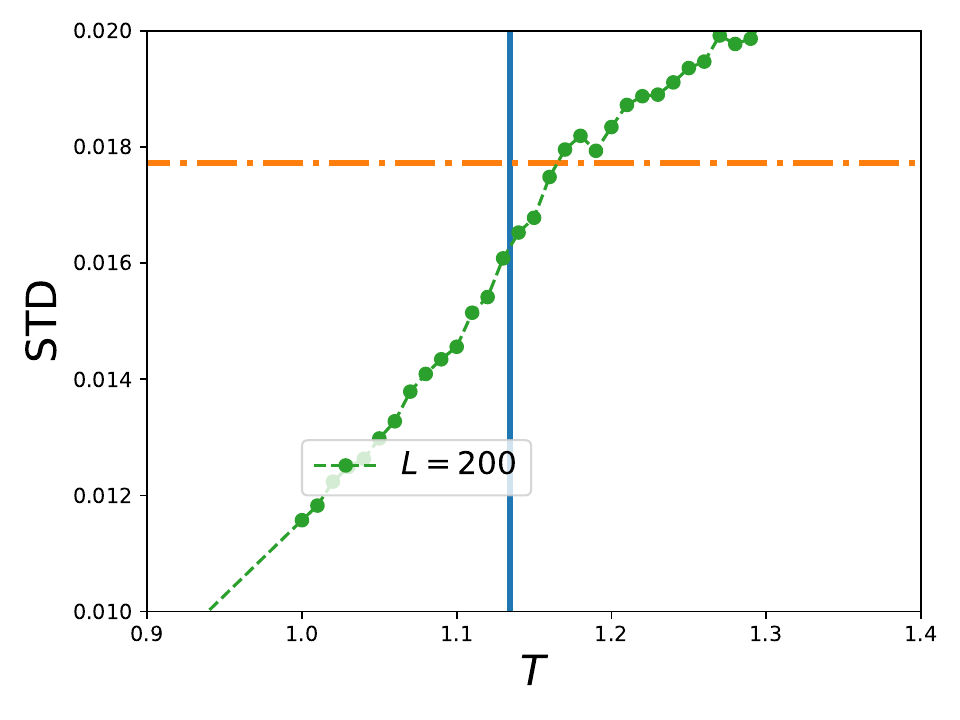}
	}
	%\vskip-0.2cm
	\caption{The intersections of $R$ and $R_{\text{H}}/2$ for $L=16$ (left panel),
		$L=32$ (middle panel), and $L=200$ (right panel).
	}
	\label{q2pseudoTc}
\end{figure*}

Fig.~\ref{q2pseudoTc} shows the intersections of $R$ curves and $R_{\text{H}}/2$ (the dashed-dotted horizontal lines) for $L=16$ (left panel), $L=32$ (middle panel), and $L=200$ (left panel).
As can be seen from the figure, $T_c(L)$ approaches the true $T_c$, which is the vertical solid
lines in the figure, as $L$ increases.

\begin{figure*}
 	%\vskip-0.5cm
 	\hbox{~~~~~~~
 	\includegraphics[width=0.4\textwidth]{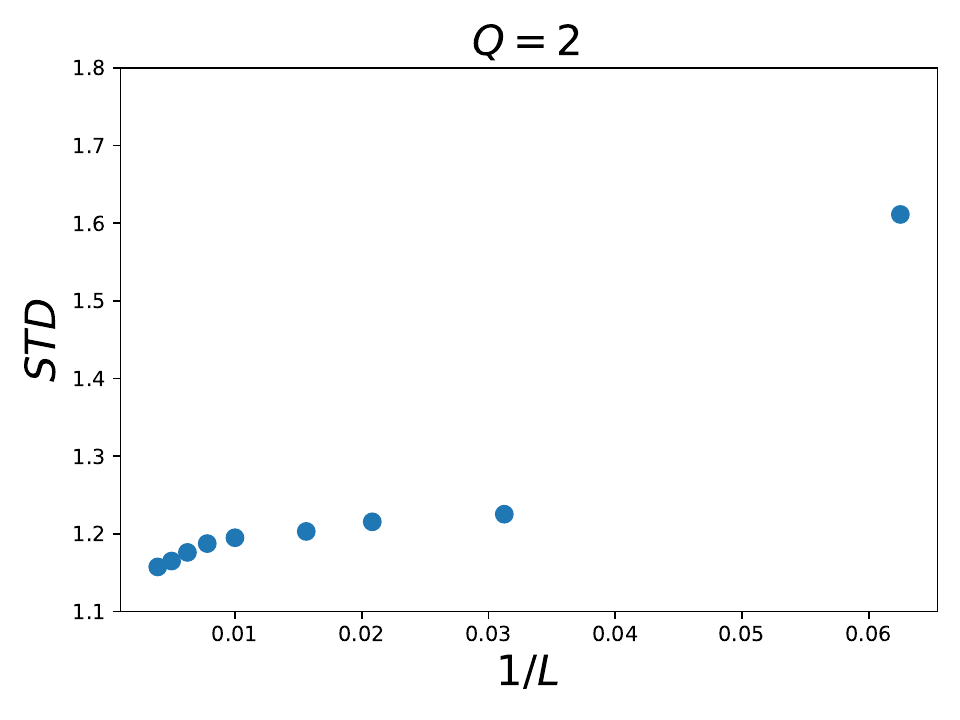}~~~~
 		\includegraphics[width=0.4\textwidth]{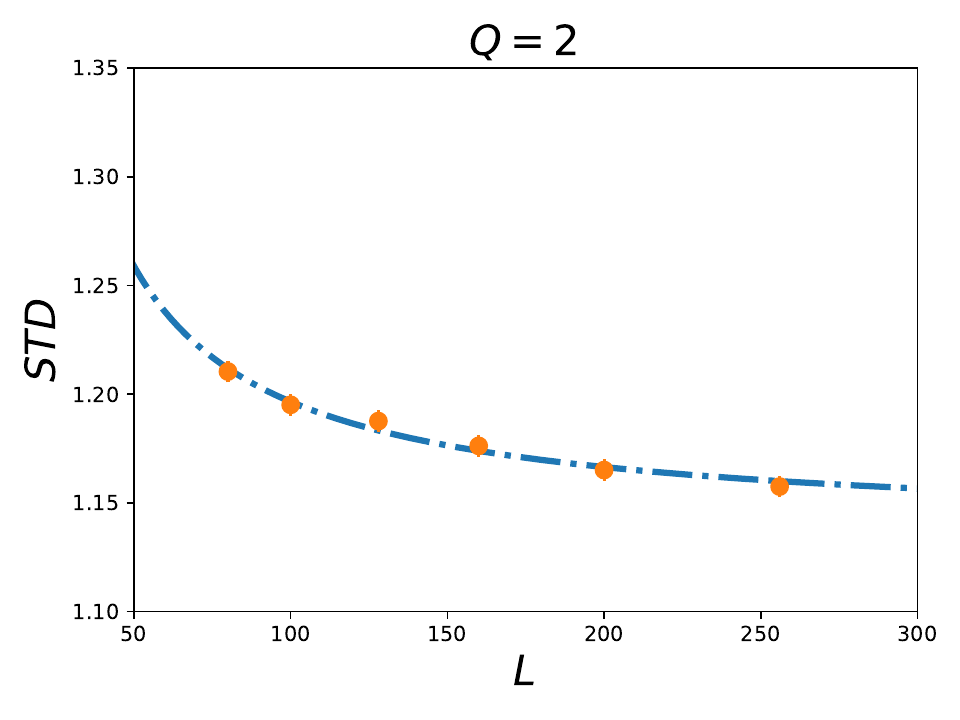}
 }
 	%\vskip-0.2cm
 	\caption{(Left) $T_c(L)$ as a function of $1/L$ for the 2-state antiferromagnetic Potts model. (Right) Fits of the data with $L \ge 80$ to the ansatz $T_c(L) = T_c \exp(-a/L)$. The solid line is obtained using the results from the fits.}
 	\label{T_cLq2}
 \end{figure*}

The pseudo-critical temperatures $T_c(L)$ as a function of $1/L$ is demonstrated in the right panel of fig.~\ref{T_cLq2}. Based on the $1/L$-behavior of $T_c(L)$ for large $L$, we fit the data of 
$80 \le L \le 256$ to the ansatz $T_c(L) = T_c\exp(-a/L)$, where $a$ is some constant. The fits lead to $T_c = 1.137(4)$
which agrees quantitatively with the expected $T_c = 1.1346$, see the left panel of fig.~\ref{T_cLq2}. The solid curve appearing in the panel is obtained using the results from
the fits.

\subsubsection{The NN outcomes of the 2D 3-state antiferromagnetic Potts model}

\begin{figure*}
	%\vskip-0.5cm
	\includegraphics[width=0.6\textwidth]{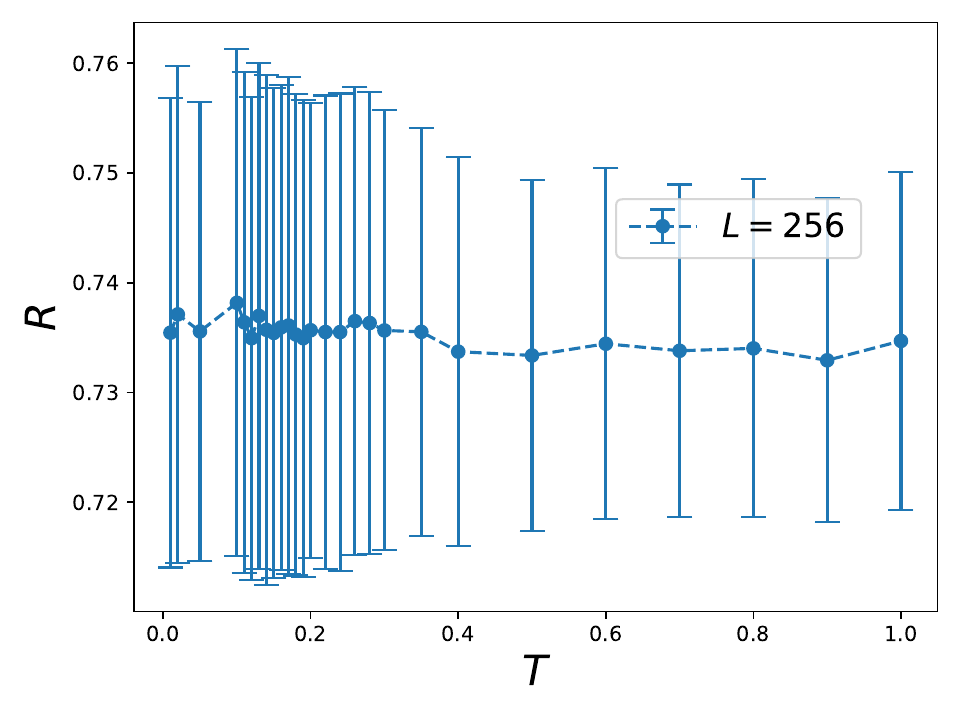}
	\vskip-0.2cm
	\caption{$R$ as a function of $T$ for $L=256$ for the 3-state antiferromagnetic Potts model.
        The results are obtained with the training set having 1000 sites.}
	\label{q3R}
\end{figure*}

$R$ as a function of $T$ for $L=256$ is demonstrated in fig.~\ref{q3R}. It is clear from
the figure that the values of $R$ are more or less the same for the considered $T$. 
For a configuration that its Potts variables are random,
it is in strong constrast to the configurations of the training set. As a result, it is anticipated that the output vector is close to (0.5,0.5) which leads to $R\sim 1/\sqrt{2}$.
Similarly, for a configuration with the antiferromagnetic ordering, there is an obvious
difference between such a configuration and that in the training set. Hence, one expects
the corresponding output vector is near (0.5,0.5) as well. With this one obtains $R \sim 1/\sqrt{2}$. This is indeed what's been observed in fig.~\ref{q3R}.

\begin{figure*}
	%\vskip-0.5cm
	\hbox{~~~~~~
		\includegraphics[width=0.3\textwidth]{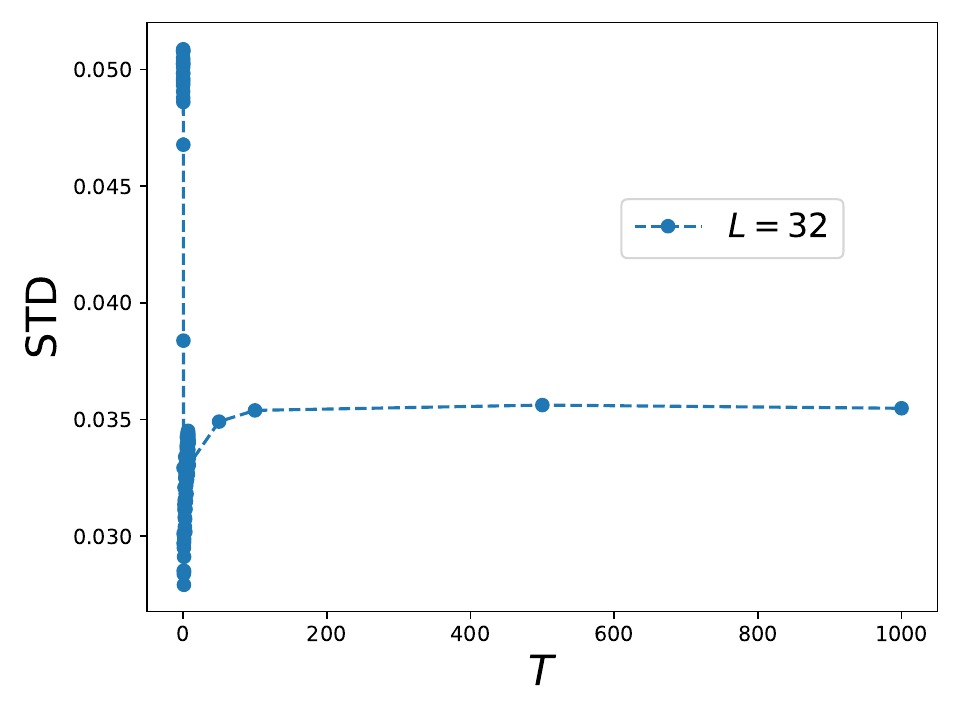}
		\includegraphics[width=0.3\textwidth]{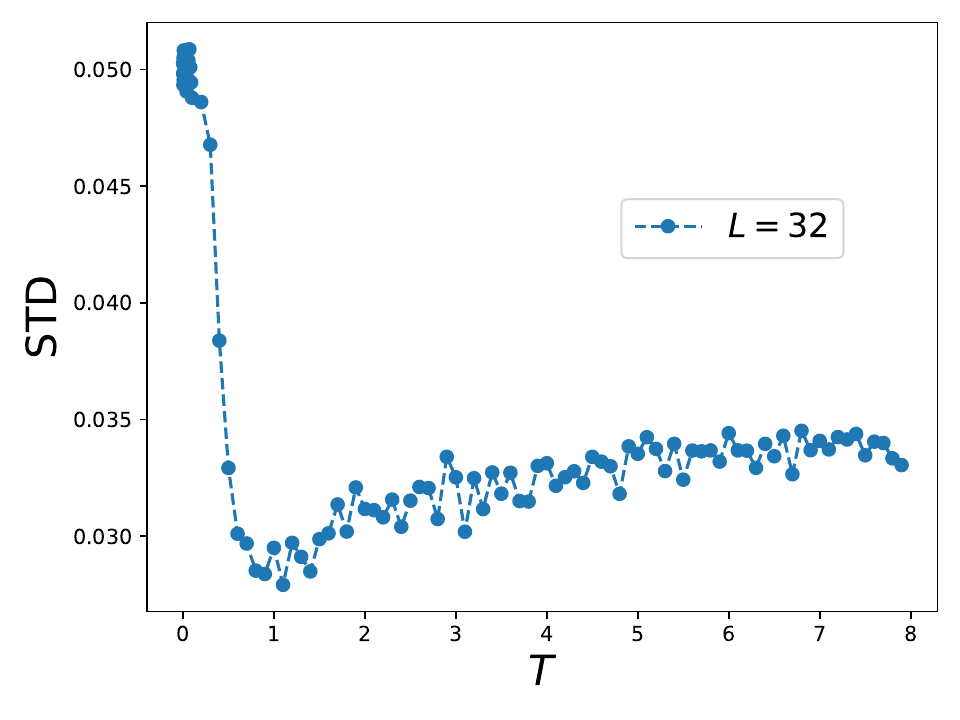}
		\includegraphics[width=0.3\textwidth]{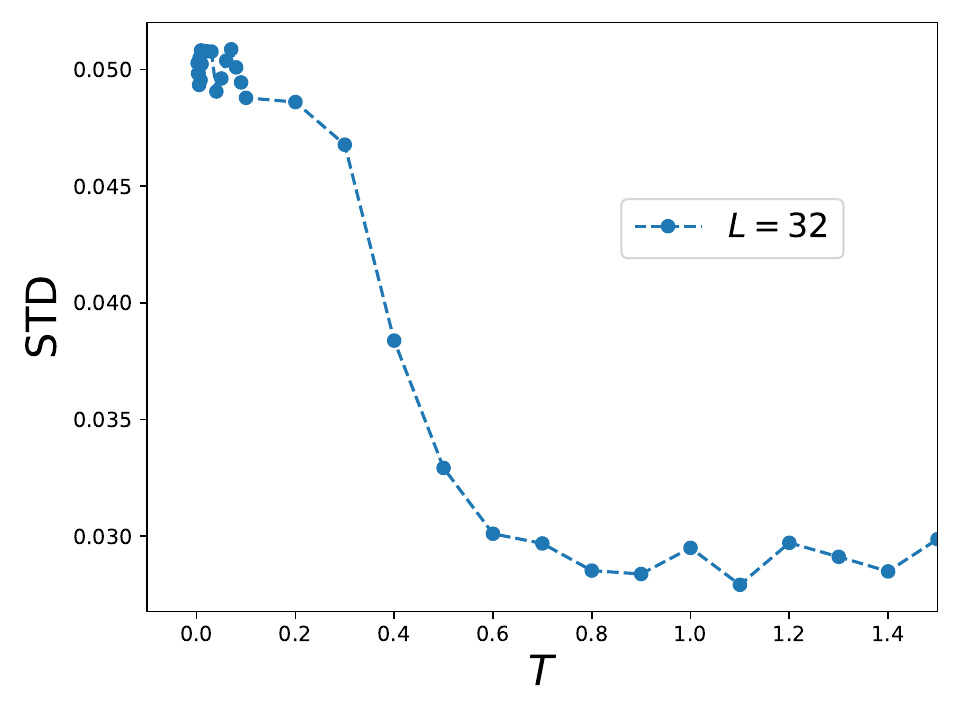}
	}
	%\vskip-0.2cm
	\caption{The STD of $R$ (left panel) as a function of $T$ for the 3-state antiferromagnetic
		Potts model for $L = 32$. The middle and the right panels are the associated STD close to $T=0$.
	}
	\label{q3STD32}
\end{figure*}

The STD of $R$ as a function of $T$ for the 3-state antiferromagnetic Potts model for $L = 32$
is shown in the left panel of fig.~\ref{q3STD32}. The middle and the right panels are the associated STD close to $T=0$. The outcomes in fig.~\ref{q3STD32} indicate that as one moves
from the high-$T$ region ($T \sim 200$) to the low-$T$ region ($T \sim 1$), STD descreases
monotonically. In addition, as $T$ is further lowered, the corresponding STD rises rapidly
around $T \sim 0.4$ and then satures to a constant when $T \le 0.2$. In particular, the saturated STD values at high- and low-temperature regions are different from each other statistically. This in turn indicates that the nature of the phase of the low-temperature 
region ($T \le$ 0.2)
may be different from that of the phase of the high-temperature region. 
The value $T$ where STD saturates is around $T \sim 0.2$ which agrees with the temperature
where the antiferromagnetic order is observed (see fig.~\ref{q3MCsnap32}).

\begin{figure*}
	%\vskip-0.5cm
	\hbox{~~~~~~
		\includegraphics[width=0.475\textwidth]{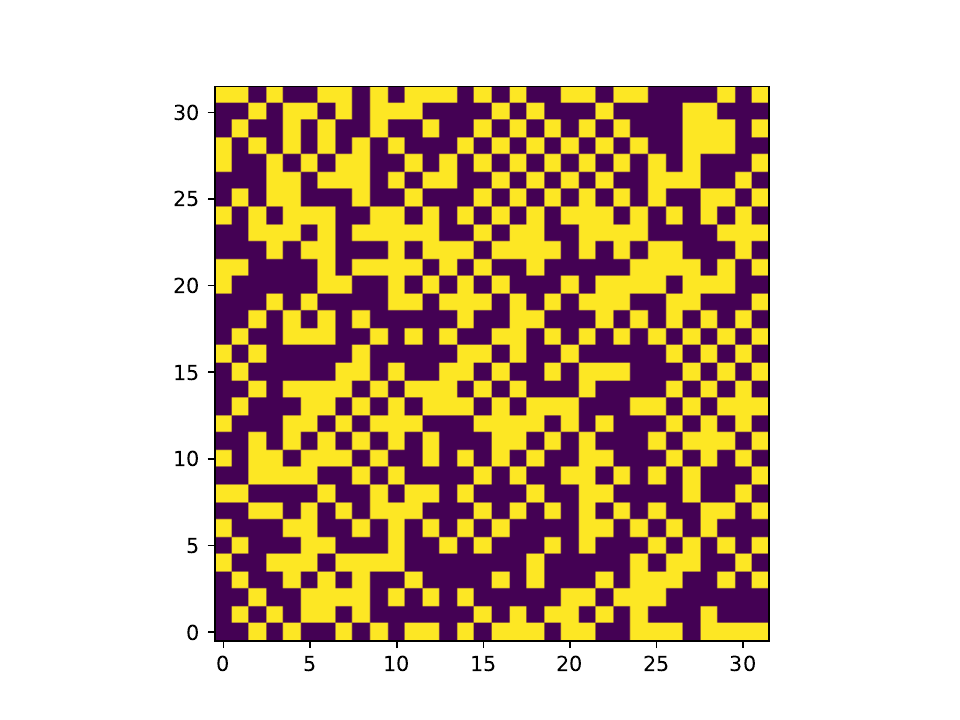}
		\includegraphics[width=0.475\textwidth]{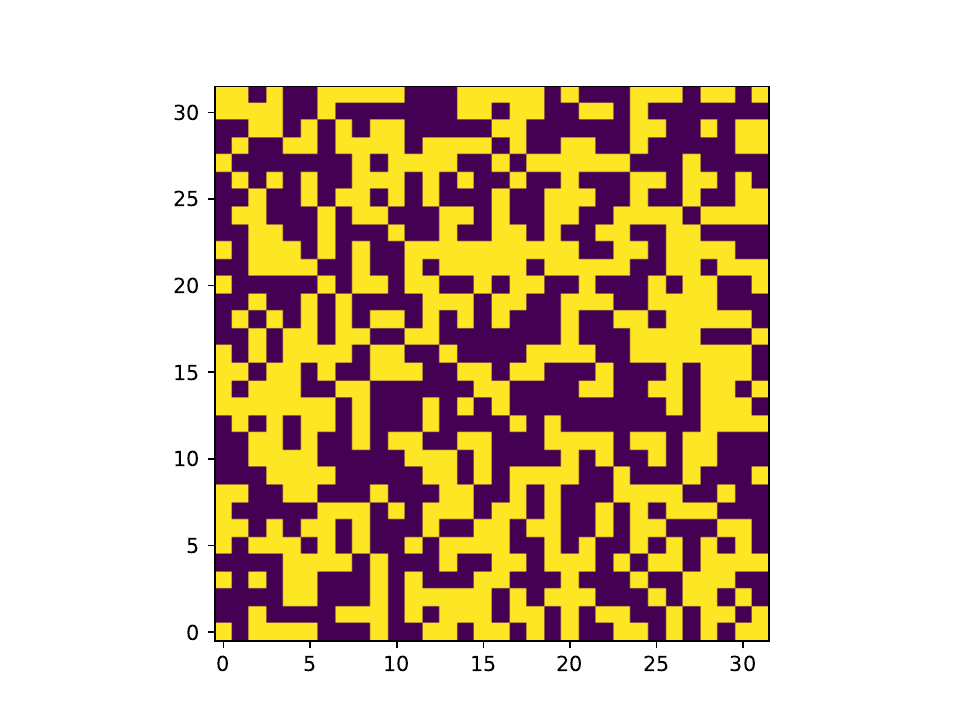}
	}
	%\vskip-0.2cm
	\caption{Typical snapshot of the configurations (of the 3-state antiferromagnetic Potts model) used for the MLP prediction. The left panel and the right panel correspond to $T=0.2$ and $T = 10.0$, respectively. The linear system sizes for both panels are $L=32$.
	}
	\label{MLPCONF32}
\end{figure*}

\begin{figure*}
	%\vskip-0.5cm
	\hbox{~~~~~~
		\includegraphics[width=0.475\textwidth]{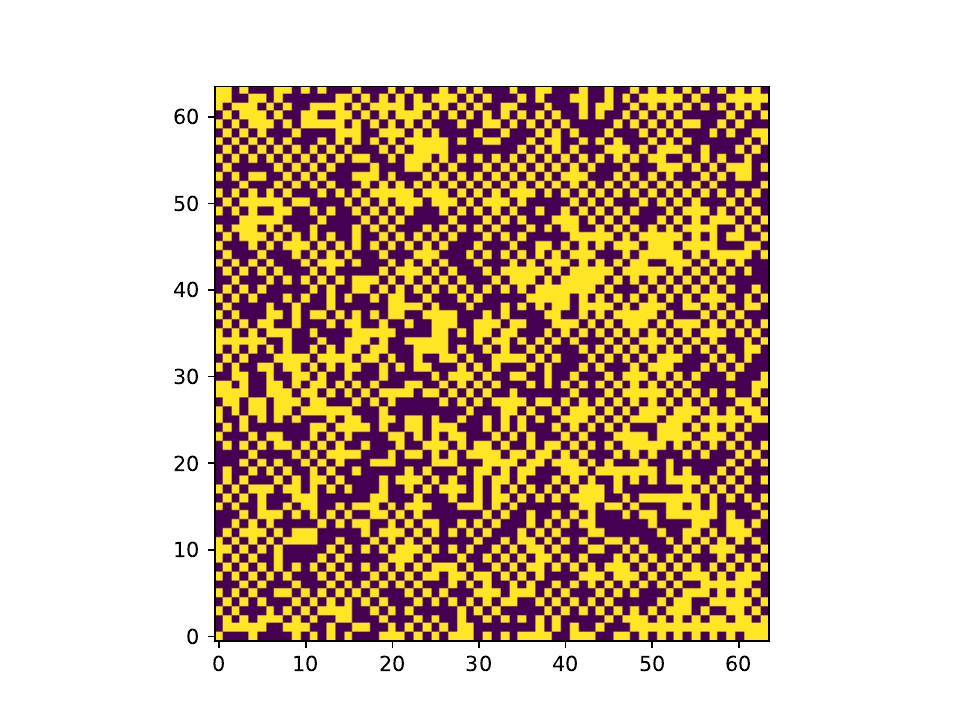}
		\includegraphics[width=0.475\textwidth]{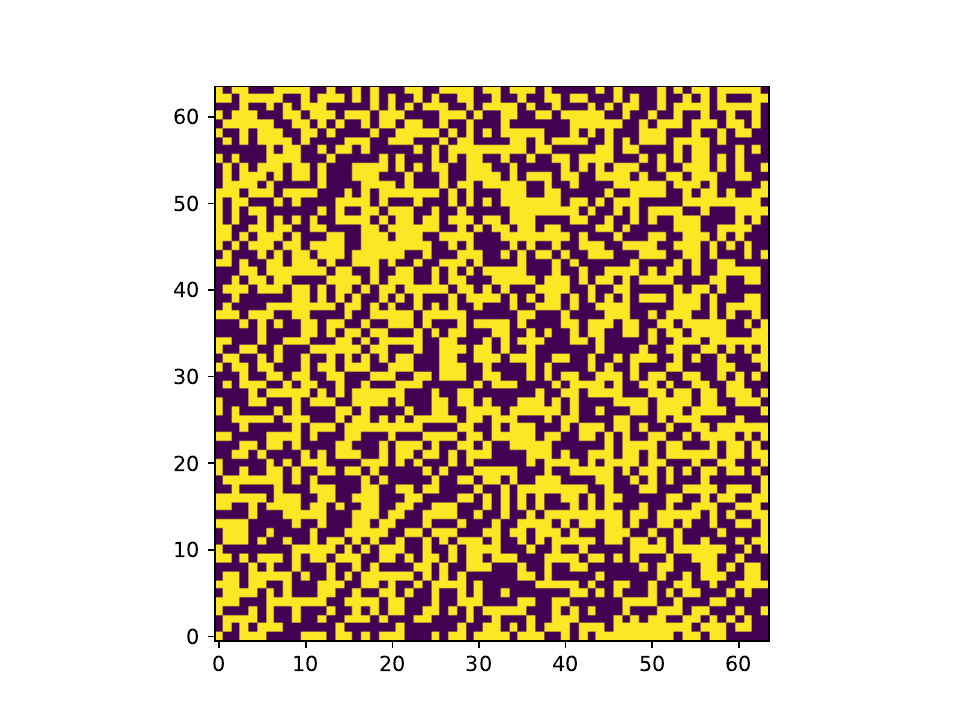}
	}
	%\vskip-0.2cm
	\caption{Typical snapshot of the configurations (of the 3-state antiferromagnetic Potts model) used for the MLP prediction. The left panel and the right panel correspond to $T=0.2$ and $T = 10.0$, respectively. The linear system sizes for both panels are $L=64$.
	}
	\label{MLPCONF64}
\end{figure*}

Interestingly, the typical configurations associated with the NN (MLP) prediction are shown in fig.~\ref{MLPCONF32}. The left and the right panels correspond to $T=0.2$ and $T=10.0$, respectively. As can be seen from the figure, the configuration at $T=0.2$ (low-temperature region) has much more isolated dark squares than that of $T=10.0$ (high-temperature region). This can be interpreted as that the configurations at low-temperature region is less random than that of
the high-temperature region. In other words, some kind of " order ", i.e. the appearance of isolated dark squares, does show up at the low-temperature region. In particular, the location and number of isolated dark squares for (any) two configurations at the low-temperature region may differ from each other significantly. 
Such a difference leads to the secanrio that the values of STD where antiferromagnetic order exists are larger than those where there is no antiferromagnetic order. This explains the result appearing in fig.~\ref{q3STD32}.  
Similar situation is found for $L=64$ as well, sees fig.~\ref{MLPCONF64}.

In conclusion, the MLP outcomes for the 3-state antiferromagnetic Potts model are consistent
with that obtained from analyzing the MC data. In particular, the temperature where
STD saturates to a constant at the low-temperature region can be considered as the temperature that the system enters the critical region.

\subsubsection{The NN outcomes for the 2D 4-state antiferromagnetic Potts model}

\begin{figure*}
	%\vskip-0.5cm
	\hbox{~~~~~~
		\includegraphics[width=0.3\textwidth]{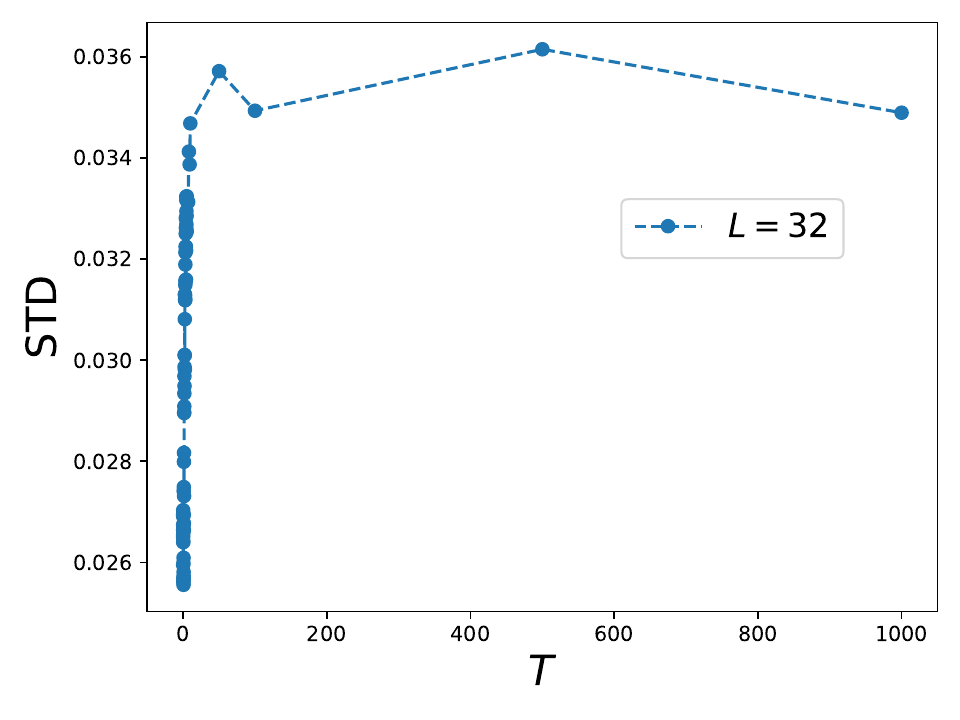}
		\includegraphics[width=0.3\textwidth]{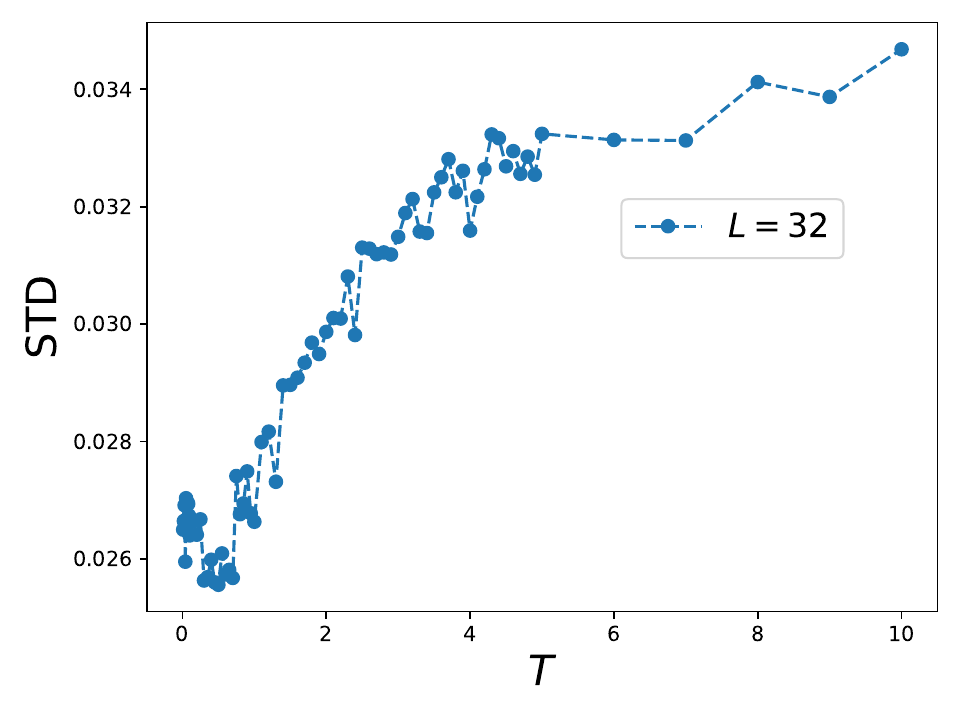}
		\includegraphics[width=0.3\textwidth]{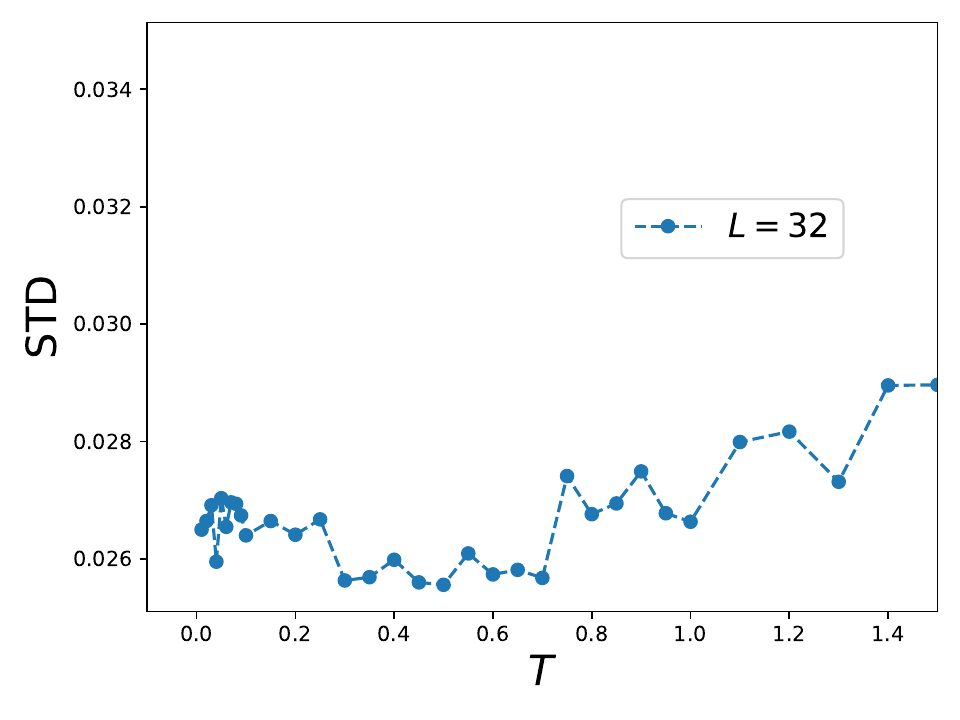}
	}
	%\vskip-0.2cm
	\caption{The STD of $R$ (left panel) as a function of $T$ for the 4-state antiferromagnetic
		Potts model for $L = 32$. The middle and the right panels are the associated STD close to $T=0$.
	}
	\label{q4STD32}
\end{figure*}

The STD of $R$ as a function of $T$ for the 4-state antiferromagnetic Potts model for $L = 32$
is shown in the left panel of fig.~\ref{q4STD32}. The middle and the right panels are the associated STD close to $T=0$. The outcomes in fig.~\ref{q4STD32} indicate that as one moves
from the high-$T$ region ($T\sim 10$) to the low-$T$ region ($T\sim0.01$), the values of STD descrease. This behavior is similar to that of $T \ge 1.0$ for the 3-state antiferromagnetic Potts model.

For the 3-state antiferromagnetic Potts model, the rising of STD around $T \le 0.4$ and the saturation of STD to a constant for $T \le 0.2$ is an indication that $T \le 0.2$ ($T_c = 0$)
is the critical region (critical temperature) for this model.
Such a phenomenon is not found for the 4-state antiferromagnetic Potts model. Based on this
as well as the description in the previous paragraph, one concludes that the 4-state antiferromagnetic Potts model is disordered for all $T$ including the zero temperature.
In other words, the MLP outcomes agree with that obtained from MC data.

\subsection{The NN results obtained from the conventional deep AECs}

\begin{figure*}
 %\vskip-0.5cm
 \includegraphics[width=0.6\textwidth]{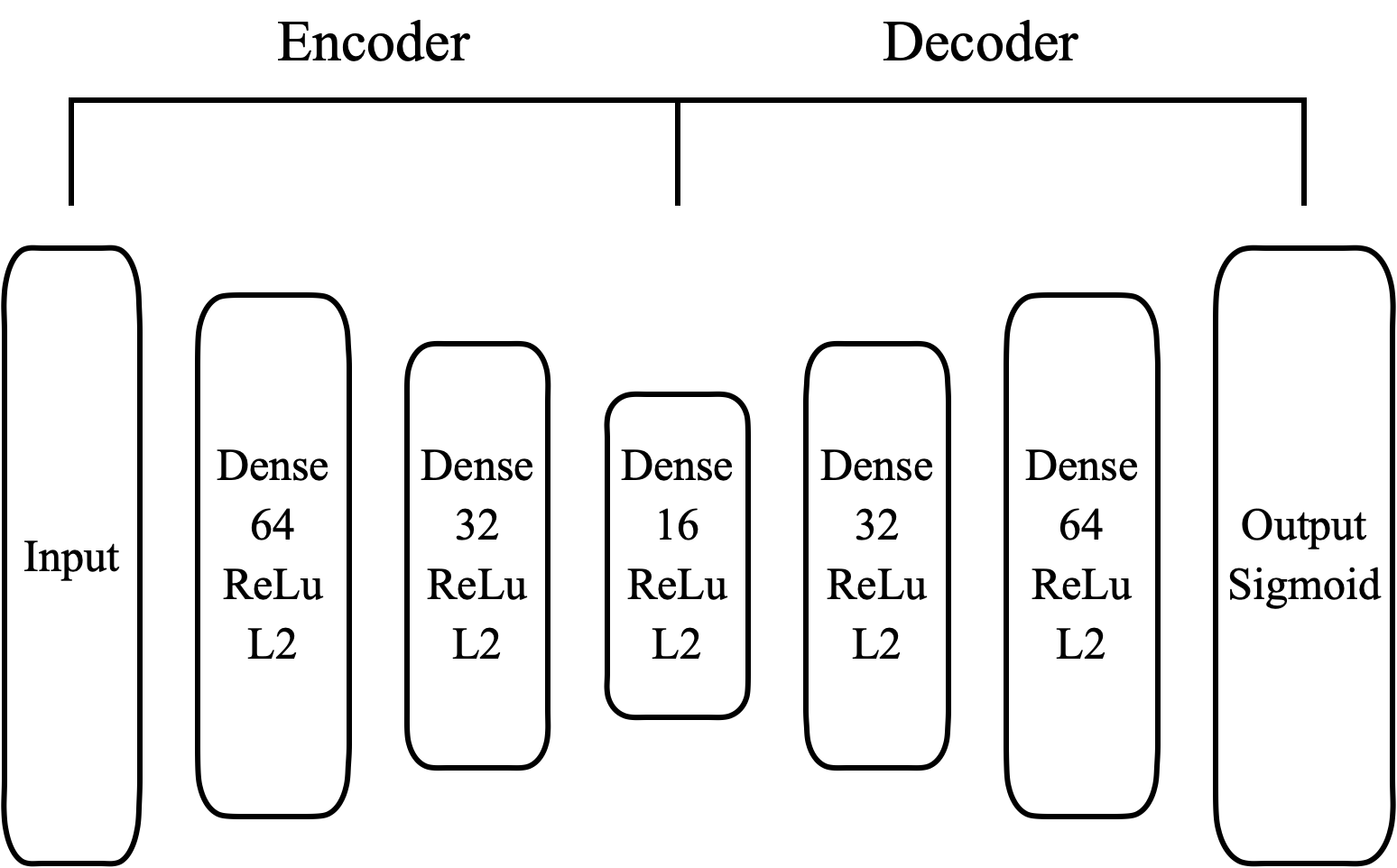}
 %\vskip-0.2cm
 \caption{The deep autoencoder employed in this study.}
 \label{ourAEC}
\end{figure*}

To examine whether the conventional unsupervised NN can determine the critical behaviors correctly
for the 2D $q$-state antiferromagnetic Potts model on the square lattice for $q=2,3,4$, we have constructed
a deep AEC consisting of one input layer, 5 hidden layers, and one output layer.
The number of neurons for
the 5 hidden layers are 64, 32, 16, 32, and 64, respectively. The AEC is trained with 800 epochs and the batchsize
used is 30. Fig.~\ref{ourAEC} is a pictorial representation of the described AEC. The used activations functions are indicated in the boxes of figure \ref{ourAEC}. $L_2$
regularization is used for each hidden layer as well.

For configurations with linear system size $L$, the outputs from the AEC are $L \times L$ matrices. In addition, the norm $R$ and the associated 
standard deviation (STD) of the output matrices are considered to determine the critical temperature $T_c$. 

\begin{figure*}
	%\vskip-0.5cm
	\hbox{~~~~~~
		\includegraphics[width=0.425\textwidth]{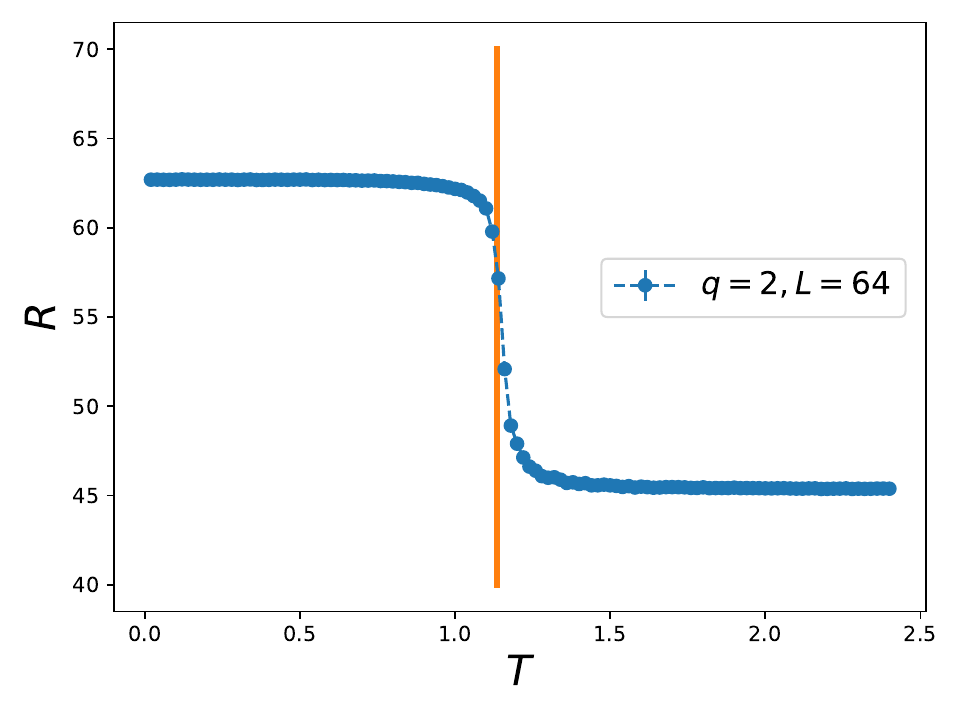}~~~~~~
		\includegraphics[width=0.425\textwidth]{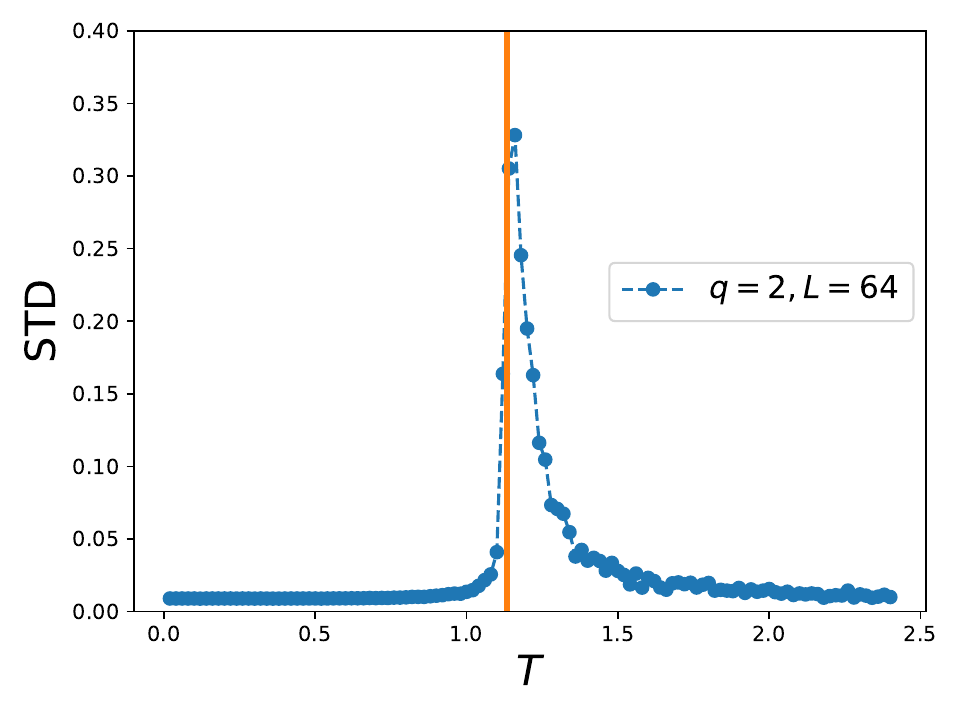}
	}
	%\vskip-0.2cm
	\caption{$R$ (left panel) and the associated STD (right panel) as functions of $T$ for the two-state antiferromagnetic Potts model on the square lattice
		with $L=64$. The vertical lines in both panels are the expected $T_c$.
	}
	\label{q2AEC}
\end{figure*}

The $R$ and the associated STD for $q=2$ and $L=64$ are shown in the left and the right panels of fig.~\ref{q2AEC}, respectively.
As can be seen from the figure, near $T_c$, there is a sudden jump (rise) in $R$ (STD). The results presented in fig.~\ref{q2AEC}
provide evidence to support the fact that the AEC is capable of determining the $T_c$ of the two-state antiferromagnetic Potts model.

After demonstrating that the constructed AEC can calculate the $T_c$ of the two-state antiferromagnetic Potts model accurately, we turn to the cases of three- and four-state antiferromagnetic Potts model.

\begin{figure*}
	%\vskip-0.5cm
	\hbox{~~~~~~
		\includegraphics[width=0.425\textwidth]{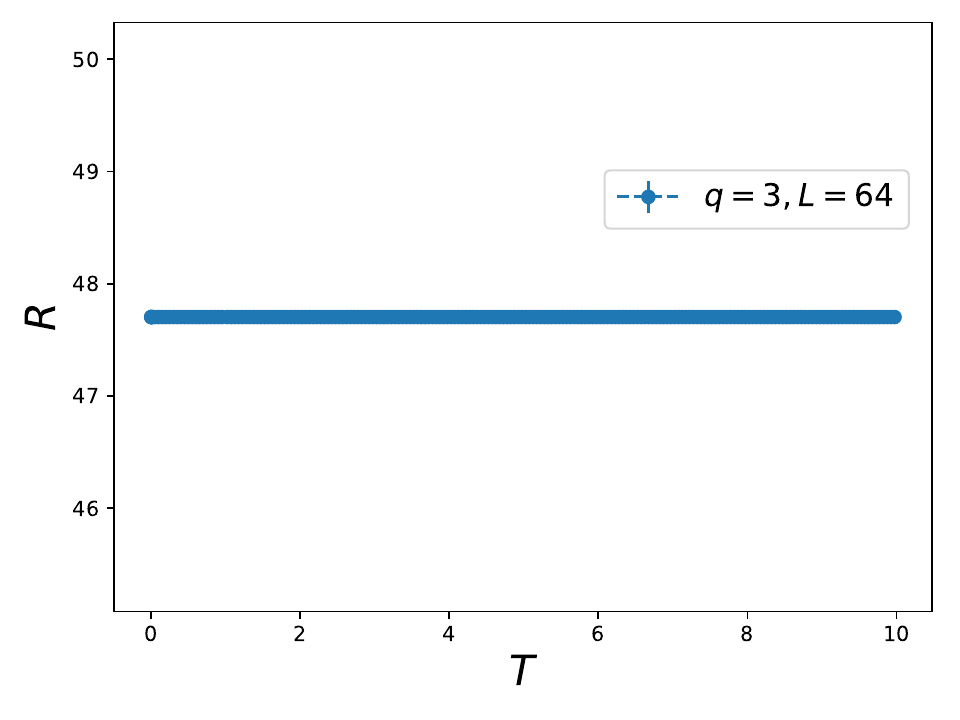}~~~~~~
		\includegraphics[width=0.425\textwidth]{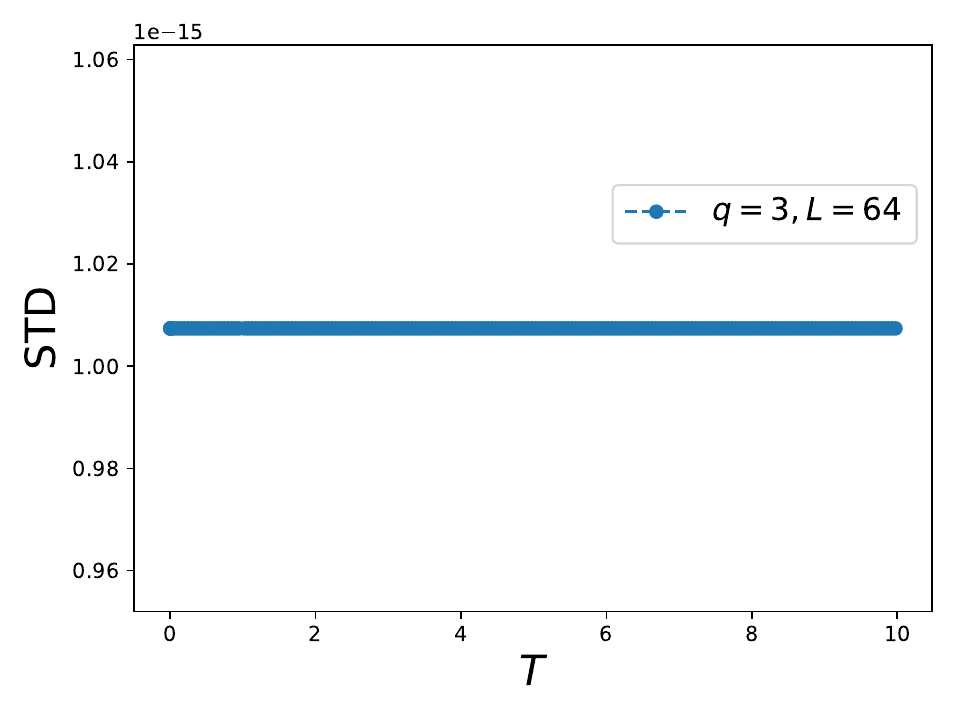}
	}
	%\vskip-0.2cm
	\caption{$R$ (left panel) and the associated STD (right panel) as functions of $T$ for 3-state antiferromagnetic Potts model on the square lattice
		with $L=64$. The results of both panels are obtained with the AEC shown in fig.~\ref{ourAEC}
	}
	\label{q3AEC}
\end{figure*}

The $R$ and the associated STD as functions of $T$ for the three-state antiferromagnetic Potts model are shown in the left and the right panels of fig.~\ref{q3AEC}. The results appearing in
the figure indicate that $R$ or the corresponding STD stays a constant for all considered
temperatures. In other words, the built AEC fails to reveal any information regarding the
critical behavior of the system.

\begin{figure*}
	%\vskip-0.5cm
	\hbox{~~~~~~
		\includegraphics[width=0.425\textwidth]{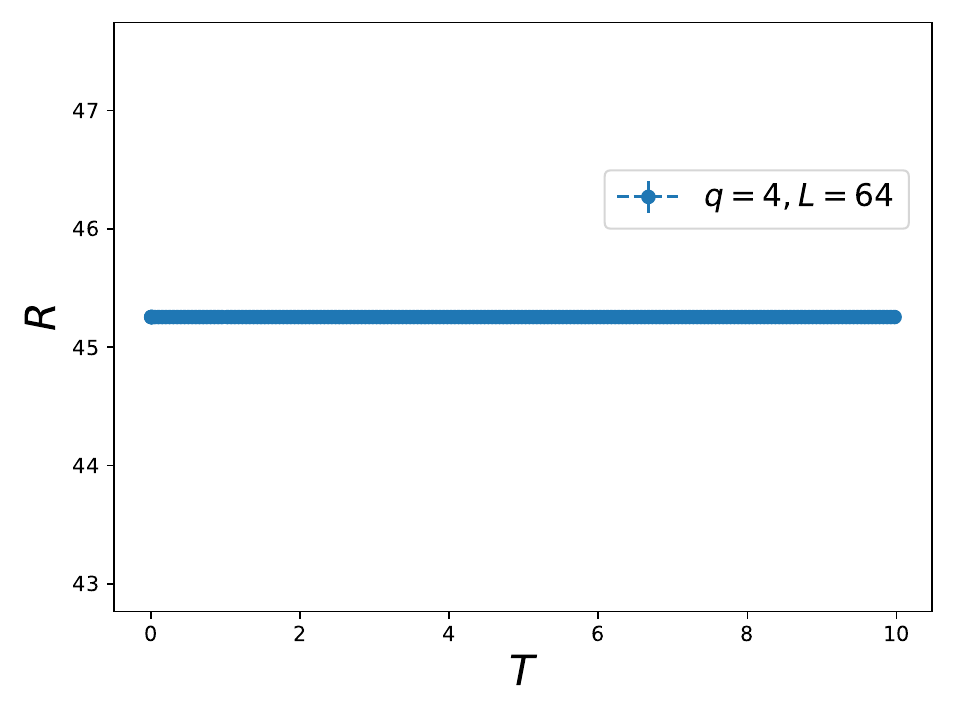}~~~~~~
		\includegraphics[width=0.425\textwidth]{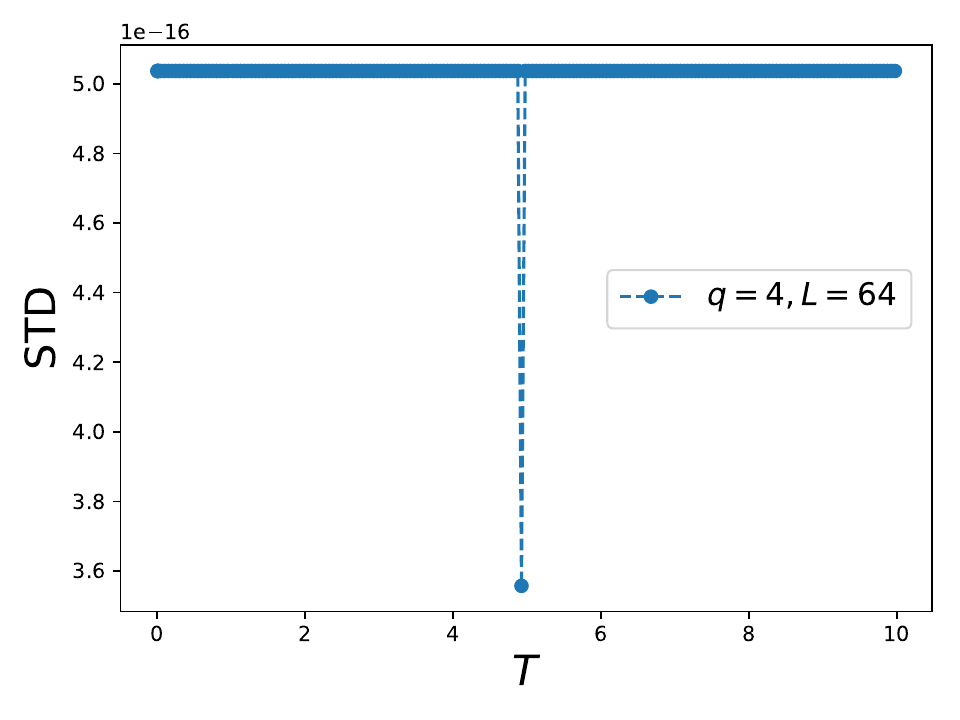}
	}
	%\vskip-0.2cm
	\caption{$R$ (left panel) and the associated STD (right panel) as functions of $T$ for 4-state antiferromagnetic Potts model on the square lattice
		with $L=64$. The results of both panels are obtained with the AEC shown in fig.~\ref{ourAEC}
	}
	\label{q4AEC}
\end{figure*}

Similarly, the conventional deep AEC cannot provide us with any relevant signal for the
critical phenomenon of the 2D 4-state antiferromagnetic Potts model, sees both panels
of fig.~\ref{q4AEC}.

\begin{figure*}
	%\vskip-0.5cm
	\includegraphics[width=0.8\textwidth]{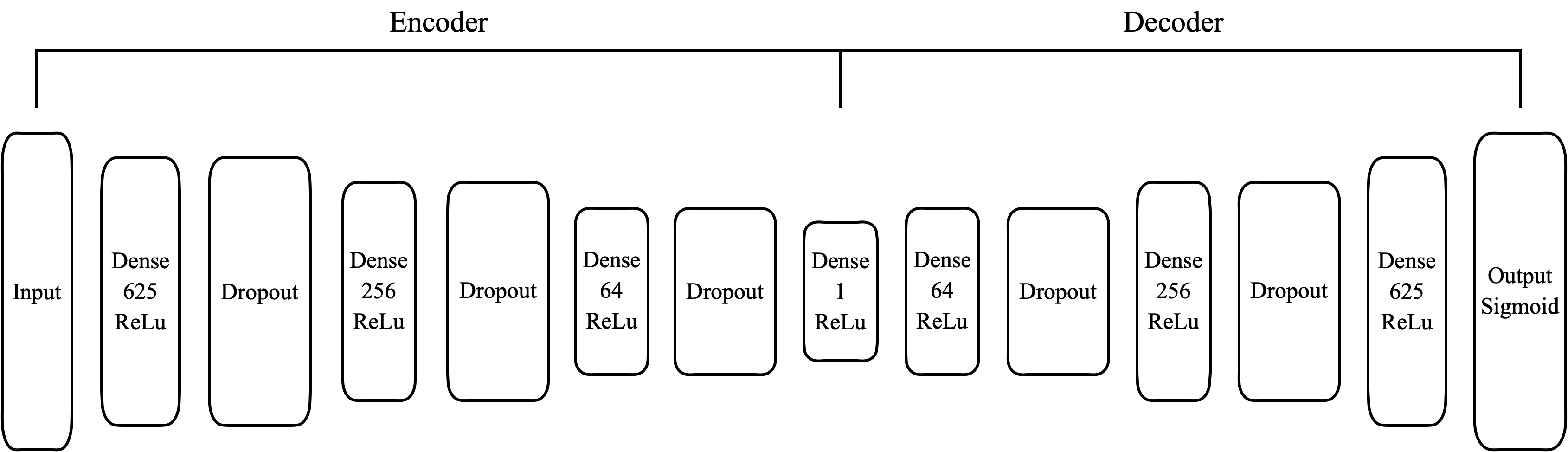}
	%\vskip-0.2cm
	\caption{The deep autoencoder of Ref.~\cite{Ale20} and considered in this study.}
	\label{cAEC}
\end{figure*}

\begin{figure*}
	%\vskip-0.5cm
	\hbox{~~~~~~
		\includegraphics[width=0.425\textwidth]{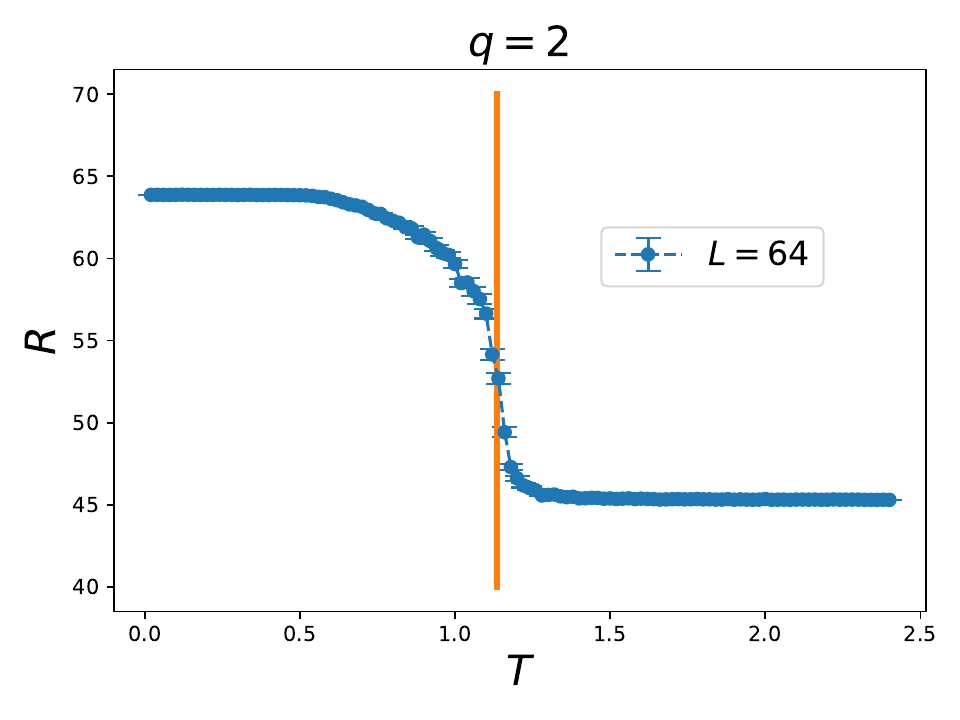}~~~~~~
		\includegraphics[width=0.425\textwidth]{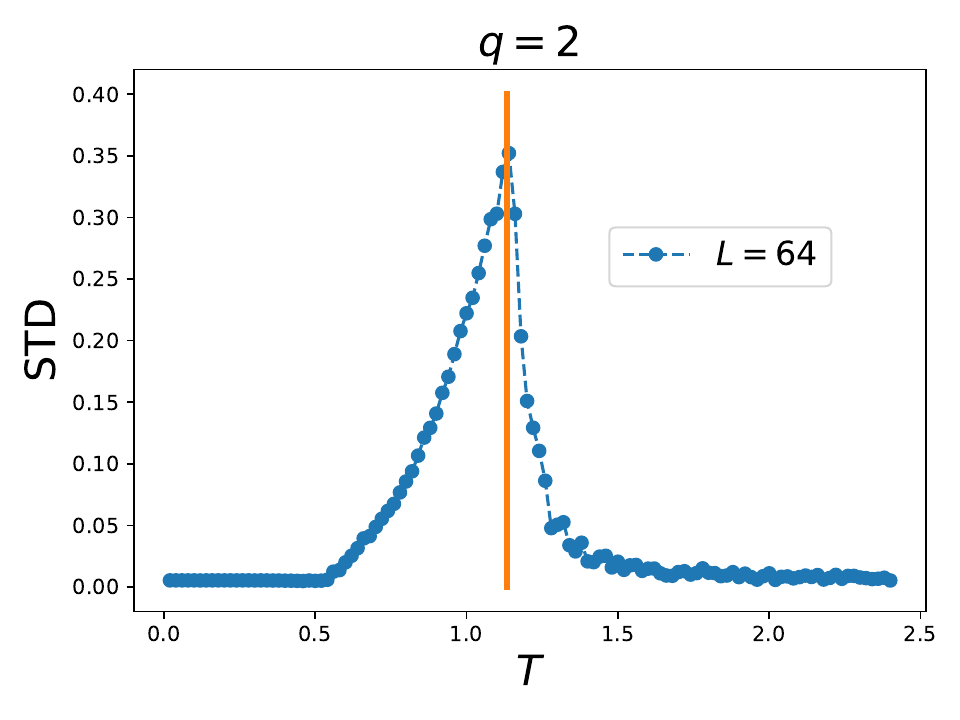}
	}
	%\vskip-0.2cm
	\caption{$R$ (left panel) and the associated STD (right panel) as functions of $T$ for the 2-state antiferromagnetic Potts model on the square lattice
		with $L=64$. The results are obtained from the AEC of Ref~\cite{Ale20}.
	}
	\label{2q2AEC}
\end{figure*}

Apart from the AEC shown in fig.~\ref{ourAEC}, we have also used a AEC with more complicated architecture \cite{Ale20} to study the crticail phenonmena
of the investigated models (see fig.~\ref{cAEC} for this mentioned
AEC). The $R$ and the associated STD for the 2-state antiferromagnetic Potts model are shown as the left and the right panel of fig.~\ref{2q2AEC}. The outcomes shown in fig.~\ref{2q2AEC} imply that the AEC of Ref.~\cite{Ale20} is capable of detecting the phase transition of the 2-state antiferromagnetic Potts model on the square lattice accurately.

\begin{figure*}
	%\vskip-0.5cm
	\hbox{~~~~~~
		\includegraphics[width=0.425\textwidth]{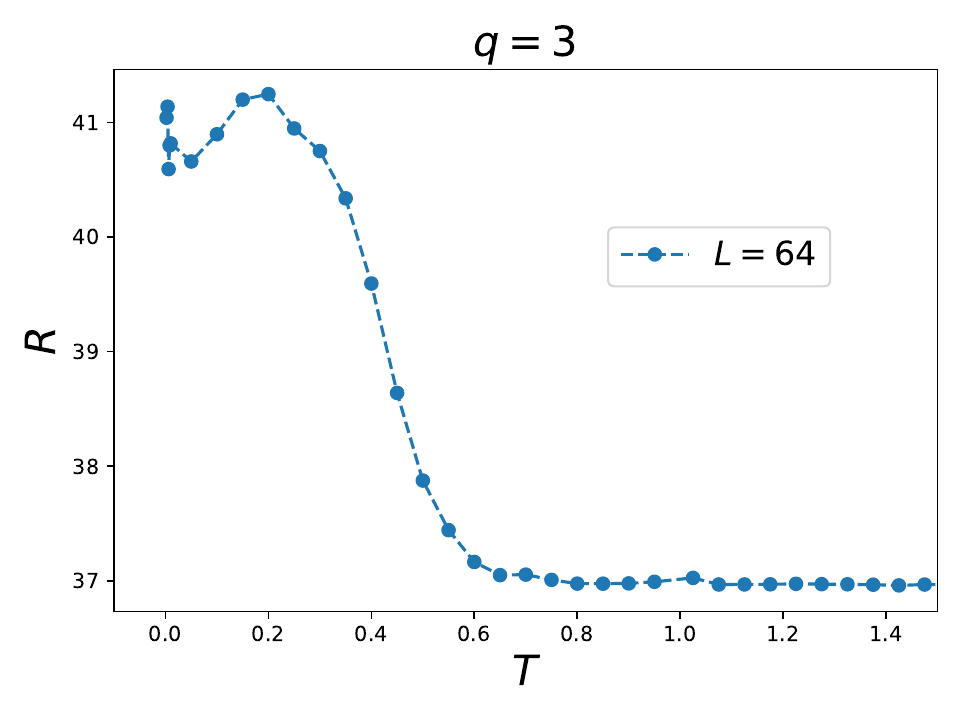}~~~~~~
		\includegraphics[width=0.425\textwidth]{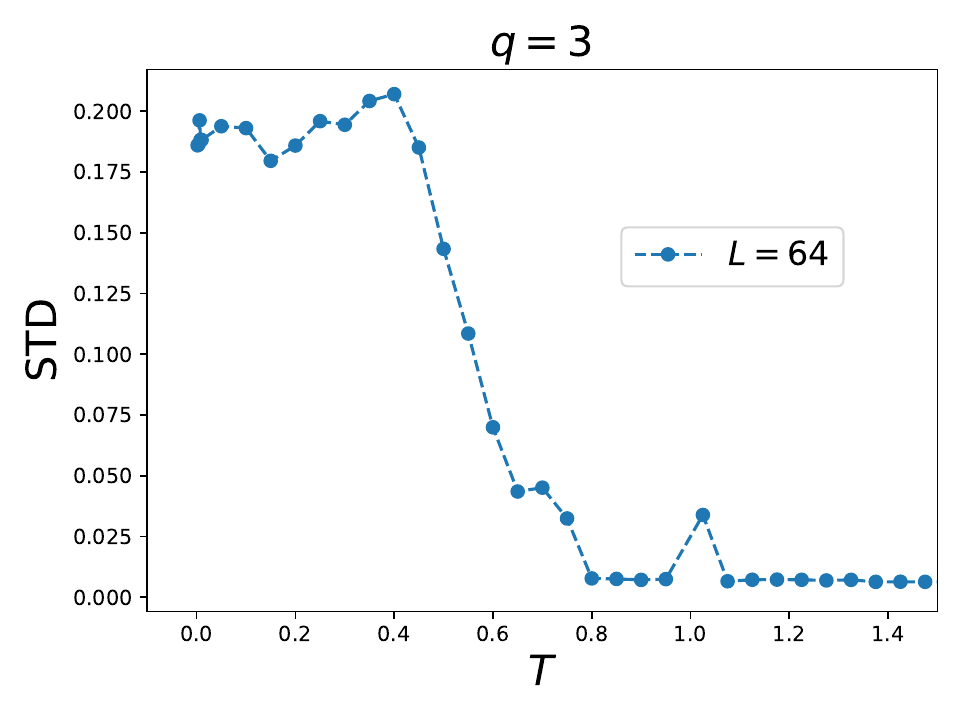}
	}
	%\vskip-0.2cm
	\caption{$R$ (left panel) and the associated STD (right panel) as functions of $T$ for the 3-state antiferromagnetic Potts model on the square lattice
		with $L=64$. The results are obtained from the AEC of Ref.~\cite{Ale20}.
	}
	\label{2q3AEC}
\end{figure*}

The $R$ and the associated STD for the 3-state antiferromagnetic Potts model, obtained from the AEC of Ref.~\cite{Ale20}, are shown in the left and the right panel of fig.~\ref{2q3AEC}. In obtaining the results, the number of configurations in the training set is half of that in the testing set. The outcomes shown in fig.~\ref{2q3AEC} imply that the AEC of Ref.~\cite{Ale20} is capable of detecting the phase transition. However, the critcal region indicates by fig.~\ref{2q3AEC} is $T \le 0.3$ (from $R$) or
$T \le 0.4$ (from STD)
which is slightly overestimated. Indeed, from the left panel of fig.~\ref{q3chimag}, the 
critical region is at least as low as $T \le 1/4 \sim 0.25$. It should be pointed out that the largest $L$ in fig.~\ref{q3chimag} is only 64. Hence one expects that the $T$ associated with the critical region is smaller than 0.25. 

In other words, while
the autoencoder of fig.~\ref{cAEC} can detect that the $T_c$ of the 3-state antiferromagnetic Potts model is small, it slightly overestimates the critical region of this model.

It is possible that with $L > 64$,
the critical region estimated by the AEC of fig.~\ref{cAEC} will be closer to the true critical region. Regarding this, we would like to emphasize the fact that with only $L=32$ data, our NN already estimates
the critical region very accurately.

\begin{figure*}
	%\vskip-0.5cm
	\hbox{~~~~~~
		\includegraphics[width=0.425\textwidth]{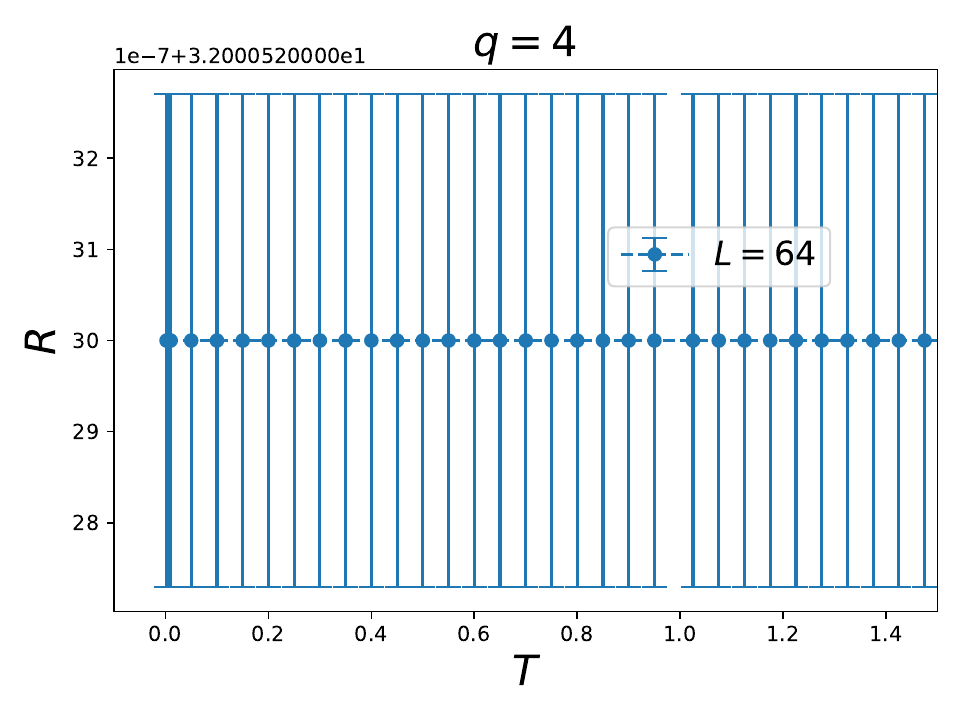}~~~~~~
		\includegraphics[width=0.425\textwidth]{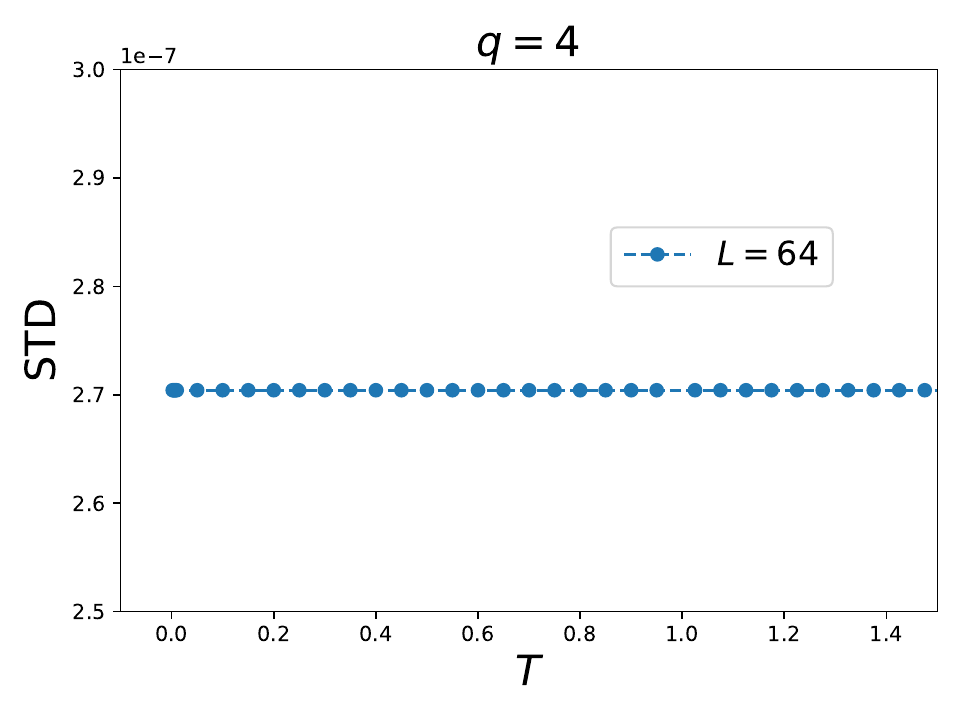}
	}
	%\vskip-0.2cm
	\caption{$R$ (left panel) and the associated STD (right panel) as functions of $T$ for the 4-state antiferromagnetic Potts model on the square lattice
		with $L=64$. The results are obtained from the AEC shown in \ref{cAEC}.
	}
	\label{2q4AEC}
\end{figure*}
 
The $R$ and the associated STD for the 4-state antiferromagnetic Potts model, obtained from the AEC of fig.~\ref{cAEC}, are shown as the left and the right panel of fig.~\ref{2q4AEC}. The outcomes shown in fig.~\ref{2q4AEC} imply that the AEC of fig.~\ref{cAEC} fails to detect correctly the critical behavior of the 4-state antiferromagnetic model model.
 
\section{Discussions and Conclusions}
 
In this study, we investigate the critical behaviors of the 2D antiferromagnetic $q$-state 
Potts model for $q=2,3,4$ using a NN trained exactly in the same way as that of Refs.~\cite{Tse22,Pen22,Tse231}. 
Only the first 400 (1000) spins are considered to construct the configurations which are then employed for the NN predictions. In particular, the standard deviations STD
of the magnitude $R$ of the output NN vectors are used as a quantity to distinguish various phases of the investigated systems.

Our NN approach successfully uncovers the
critical behaviors of the studied models. For instance, the $T_c = 1.137(4)$ of the 2-state antiferromagnetic Potts model determined by our NN agrees well with the theoretical expectation $T_c = 1.1346$. In addition, we also provide convincing evidence to show that the 3-state antiferromagnetic Potts model has $T_c = 0$ (and $T \le 0.2$ is the critical region), and the 4-state antiferromangetic Potts model
is always disordered for all temperatures including the zero temperature.
 
We also apply the conventional deep AECs including the one of Ref.~\cite{Ale20} to study the targeted critical behaviors. 
The results
suggest that although these conventioanl AECs can detect the $T_c$ of the 2-state antiferromagnetic model, they fail to give definite conclusions regarding the critical behaviors of 3- and 4-state antiferromagnetic Potts model on the square lattice.

\begin{figure*}
	%\vskip-0.5cm
	\hbox{~~~~~~
		\includegraphics[width=0.425\textwidth]{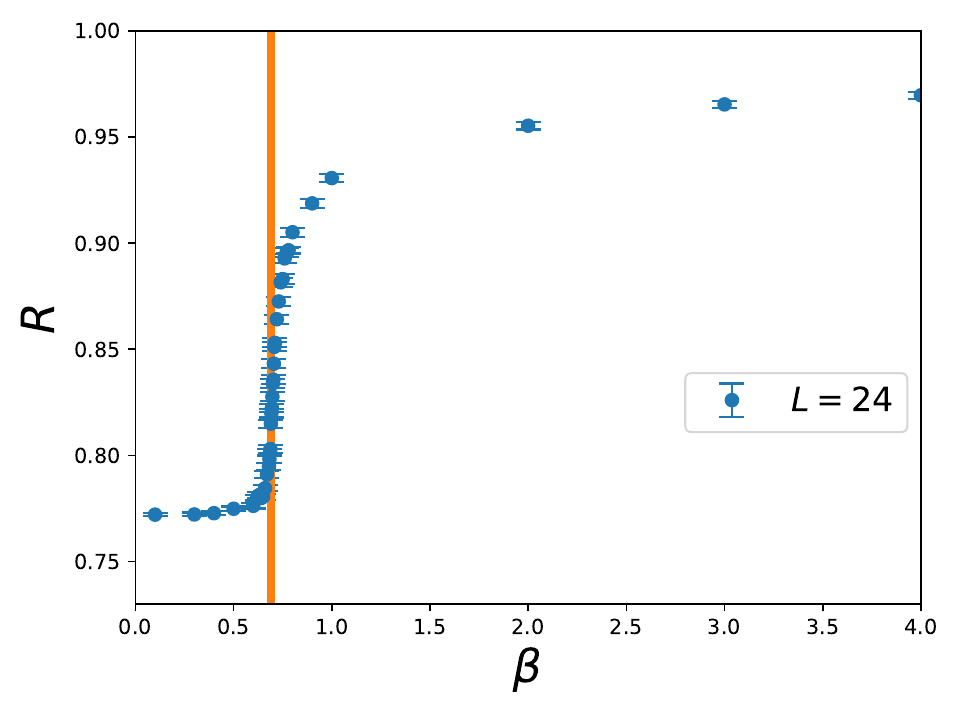}~~~~~~
		\includegraphics[width=0.425\textwidth]{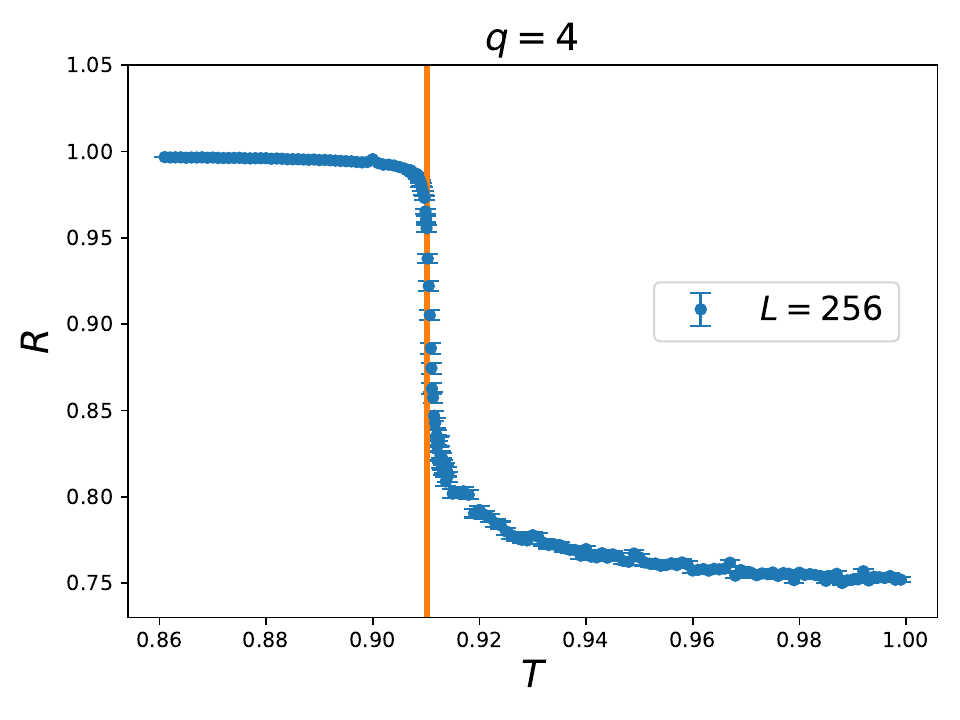}
	}
	%\vskip-0.2cm
	\caption{(left panel) $R$ as a function of the inverse temperature $\beta$ for the 3D classical $O(3)$ model (obtained with the training set having 400 sites). (Right panel) $R$ as a function of $T$ for the 4-state ferromagnetic Potts model on the square lattice (obtained with the training set having 1000 sites).}
	\label{o3q4}
\end{figure*}

Previously, our NN has been applied to study the critical theories of many 3D and 2D models
including the 3D $O(3)$ model, the 2D $XY$ model, the 2D generalized 3-state $XY$ model,
the 2D classical 6-state and 8-state clock models, 2D $U(1)$ quantum link model, and the 2D ferromagnetic $q$-state Potts models with $q=2,3,4,6,8,9,10$. Here a variant of our NN is employed to investigate the 2D antiferromagnetic $q$-state Potts model for $q=2,3,4$. 
An interesting problem to explore is to examine whether the method of constructing
the needed configurations for the NN prediction used in this study is also valid for
the models mentioned earlier in this paragraph. The left and the right panels of fig.~\ref{o3q4} show $R$ as functions of the inverse temperature $\beta$ and $T$
for the 3D classical $O(3)$ model and 2D 4-state ferromagnetic Potts model on the square lattice, respectively. The figures are obtained exactly by the approach used in this study. The vertical solid lines in both panels of fig.~\ref{o3q4} are the expected critical points. The results demonstrated in fig.~\ref{o3q4}, although seem slightly less accurate
than that shown in Ref.~\cite{Tse22,Tse231}, provide evidence to support the fact that the idea of building
configurations for the NN prediction employed here, i.e. using only the first 400 (1000) spins to construct the needed configurations for the NN prediction, also works for other models including
those referred in this paragraph.

It is interesting that a simple NN and its variants, which are trained using two artificially made configurations as the training set (no real spin configuraions are employed to train our NNs), can successfully calculate the $T_c$ of so many models
that are different dramatically from each other.

%\vskip-0.5cm
\section*{Acknowledgement}%\vskip-0.3cm

Partial support from the National Science and Technology Council (NSTC) of Taiwan is
acknowledged (NSTC 112-2112-M-003-016- and NSTC 113-2112-M-003-014-).


\begin{thebibliography}{1}


  



\bibitem{Oht16}
Tomoki Ohtsuki and Tomi Ohtsuki, {\it Deep Learning the Quantum Phase Transitions in Random Two-Dimensional Electron Systems},
J. Phys. Soc. Jpn. 85, 123706 (2016).


\bibitem{Tan16}
  Akinori Tanaka, Akio Tomiya,
{\it Detection of Phase Transition via Convolutional Neural Networks},
  J. Phys. Soc. Jpn. 86, 063001 (2017).

  
\bibitem{Car16}
  Juan~Carrasquilla, Roger~G.~Melko,
  {\it Machine learning phases of matter},
  Nature Physics {\bf 13}, 431–434 (2017). (2017). https://doi.org/10.1038/nphys4035.


\bibitem{Nie16}
Evert P.L. van Nieuwenburg, Ye-Hua Liu, Sebastian D. Huber,
{\it Learning phase transitions by confusion},
Nature Physics {\bf 13}, 435–439 (2017). https://doi.org/10.1038/nphys4037.



\bibitem{Den17}
Dong-Ling Deng, Xiaopeng Li, and S. Das Sarma,
{\it Machine learning topological states},
Phys. Rev. B {\bf 96} 195145 (2017), 1-11. https://doi.org/10.1103/PhysRevB.96.195145.

\bibitem{Li18}
  C.-D. Li, D.-R. Tan, and F.-J. Jiang,
{\it Applications of neural networks to the studies of phase transitions of two-dimensional Potts models},
  Annals of Physics, 391 (2018) 312-331. https://doi.org/10.1016/j.aop.2018.02.018.

\bibitem{Chn18}
Kelvin Ch'ng, Nick Vazquez, and Ehsan Khatami,
{\it Unsupervised machine learning account of magnetic transitions in the Hubbard model},
Phys. Rev. E {\bf 97}, 013306 (2018), 1-10. https://doi.org/10.1103/PhysRevE.97.013306.



\bibitem{Don19}
Xiao-Yu Dong, Frank Pollmann, and Xue-Feng Zhang,
{\it Machine learning of quantum phase transitions},
Phys. Rev. B {\bf 99}, 121104(R) (2019).

  
  \bibitem{Zha19}
Wanzhou Zhang, Jiayu Liu, and Tzu-Chieh Wei,
{\it Machine learning of phase transitions in the percolation and {\text{XY}} models},
Phys. Rev. E {\bf 99}, 032142 (2019), 1-14. https://doi.org/10.1103/PhysRevE.99.032142.


\bibitem{Tan20.1}
D.-R. Tan { \it et al.}
{\it A comprehensive neural networks study of the phase
	transitions of Potts model}, 2020 New J. Phys. 22 063016, 1-17. https://doi.org/10.1088/1367-2630/ab8ab410.1088/1367-2630/ab8ab4.


\bibitem{Ale20}
Constantia Alexandrou, Andreas Athenodorou, Charalambos Chrysostomou, Srijit Paul,
Eur. Phys. J. B (2020) 93: 226.
%\doi{10.1140/epjb/e2020-100506-5}


\bibitem{Fuk21}
Kimihiko Fukushima and Kazumitsu Sakai, {\it Can a CNN trained on th Ising model
	detect the phase transition of the $q$-state Potts model?},
Prog. Theor, Exp. Phys. {\bf 2021}, 061A01.


\bibitem{Tol23}
D.~W.~Tola and M. Bekele, 
{\it Machine Learning of Nonequilibrium
	Phase Transition in an Ising Model on
	Square Lattice}.
Condens. Matter 2023, 8(3), 83. https://doi.org/10.3390/condmat8030083.

\bibitem{Meh19}
P. Mehta, M. Bukov, C.-H. Wang, A. G.R. Day, C. Richardson, C. K. Fisher and D. J. Schwab,
{\it A high-bias, low-variance introduction to machine learning for physicists}, Phys. Rep. {\bf 810},
1 (2019). https://dor.org/10.1016/j.physrep.2019.03.001.




\bibitem{Car19}
G. Carleo, I. Cirac, K. Cranmer, L. Daudet, M. Schuld, N. Tishby, L. Vogt-Maranto and
L. Zdeborov\'a, {\it Machine learning and the physical sciences}, Rev. Mod. Phys. {\bf 91}, 045002
(2019). https://doi.org/10.1103/RevModPhys.91.045002.



\bibitem{Wu82}
F.~Y.~Wu, {\it The Potts model}, Rev. Mod. Phys. {\bf 54}, 235 (1982).

\bibitem{Wan89}
J.-S. Wang, R. H. Swendsen, and R. Kotecky,
Phys. Rev. Lett. {\bf 63}, 109 (1989).

\bibitem{Wan90}
Jian-Sheng Wang, Robert H. Swendsen, and Roman Kotecky,
Phys. Rev. B {\bf 42}, 2465 (1990).

\bibitem{Fer99}
Ferreira, S.J. and Sokal, A.D., 
Journal of Statistical Physics (1999) 96: 461.


\bibitem{Tse22}
Yuan-Heng Tseng, Fu-Jiun Jiang, and C.-Y. Huang,
{\it A universal training scheme and the resulting
	universality for machine learning phases},
Prog. Theor. Exp. Phys. {\bf 2023} 013A03, 1-15. https://doi.org/10.1093/ptep/ptac173.

\bibitem{Pen22}
Jhao-Hong Peng, Yuan-Heng Tseng, Fu-Jiun Jiang, 
{\it Machine learning phases of an Abelian gauge theory},
Prog. Theor. Exp. Phys. {\bf 2023} 073A03.
https://doi.org/10.1093/ptep/ptad096.

\bibitem{Tse23}
Yuan-Heng Tseng, Fu-Jiun Jiang, {\it Detection of Berezinskii–
	Kosterlitz--Thouless transitions for the two-dimensional q-state clock models
	with neural networks}, Eur. Phys. J. Plus, (2023) {\bf 138}:1118.
https://doi.org/10.1140/epjp/s13360-023-04741-4.

\bibitem{Tse231}

Yuan-Heng Tseng, and Fu-Jiun Jiang, {\it Learning the phase transitions of two-dimensional Potts model with a pre-trained one-dimensional neural network}, Results in Physics 56 (2024) 107264. https://doi.org/10.1016/j.rinp.2023.107264.

\bibitem{keras}
https://keras.io

\bibitem{tens}
https://www.tensorflow.org


\bibitem{Jia24}
Fu-Jiun Jiang, a preprint submitted to PTEP.













\end{thebibliography}
\end{document}